\documentclass[zpreprint,zbstnp]{zeus_paper}
\usepackage{lineno}
\usepackage[english]{babel}
\chardef\usc=95
\chardef\til=126
\catcode`\@=11 
\DeclareRobustCommand\xdotspace{\futurelet\@let@token\@xdotspace}
\def\@xdotspace{%
  \ifx\@let@token.\else
  \ifx\@let@token\bgroup.\else
  \ifx\@let@token\egroup.\else
  \ifx\@let@token\/.\else
  \ifx\@let@token\ .\else
  \ifx\@let@token~.\else
  \ifx\@let@token!.\else
  \ifx\@let@token,.\else
  \ifx\@let@token:.\else
  \ifx\@let@token;.\else
  \ifx\@let@token?.\else
  \ifx\@let@token/.\else
  \ifx\@let@token'.\else
  \ifx\@let@token).\else
  \ifx\@let@token-.\else
  \ifx\@let@token\@xobeysp.\else
  \ifx\@let@token\space.\else
  \ifx\@let@token\@sptoken.\else
   .\space
   \fi\fi\fi\fi\fi\fi\fi\fi\fi\fi\fi\fi\fi\fi\fi\fi\fi\fi}
\catcode`\@=12 

\newcommand{\stru}[2]{%
   \relax\ifmmode\hbox{\vrule height#1 depth#2 width0pt}%
   \else\vrule height#1 depth#2 width0pt\fi}

\newcommand{\Ronum}[1]{\uppercase\expandafter{\romannumeral#1}}
\newcommand{\ronum}[1]{\expandafter{\romannumeral#1}}
\DeclareRobustCommand{\LaTeXZ}{%
  \LaTeX\kern-.05em4\kern-.1em
  {\raisebox{-0.2ex}{$\scriptstyle\text{ZEUS}$}}\xspace}

\DeclareMathAlphabet{\mathbf}{OT1}{cmr}{bx}{sl}
\newcommand{\eVdist}{\kern-0.06667em}

\newcommand{\gev}{{\,\text{Ge}\eVdist\text{V\/}}}

\newcommand{\Tesla}{\,\text{T}}

\newcommand{\slashfrac}[2]{%
  \raisebox{0.5ex}{\ensuremath #1}\kern-0.12em/\kern-0.08em
  \raisebox{-.8ex}{\ensuremath #2}}

\newcommand{\sqr}[3]{%
    {\vcenter{\hrule height.#3ex\hbox{\vrule width.#2ex height#1ex
     \kern#1ex\vrule width.#3ex}\hrule height.#2ex}}}

\catcode`\@=11 
\newcommand{\parenbar}{\mathpalette\p@renb@r}
\def\p@renb@r#1#2{\vbox{%
  \ifx#1\scriptscriptstyle \dimen@.7em\dimen@ii.2em\else
  \ifx#1\scriptstyle \dimen@.8em\dimen@ii.25em\else
  \dimen@1em\dimen@ii.4em\fi\fi \offinterlineskip
  \ialign{\hfill##\hfill\cr
    \vbox{\hrule width\dimen@ii}\cr
    \noalign{\vskip-.3ex}%
    \hbox to\dimen@{$\mathchar300\hfil\mathchar301$}\cr
    \noalign{\vskip-.3ex}%
    $#1#2$\cr}}}
\catcode`\@=12 

\newcommand{\MSbar}{\hbox{$\overline{\rm MS}$}\xspace}

\newcommand{\IP}{{\rm I$\kern-0.01667em$P}\xspace}

\mathchardef\qsm=63
\mathchardef\pls=43
\mathchardef\mns=512
\mathchardef\plm=518
\mathchardef\eql=61
\mathchardef\smallleft=300
\mathchardef\smallright=301
\mathchardef\les=316
\mathchardef\gre=318
\mathchardef\leq=532
\mathchardef\grq=533

\catcode`\@=11 
\newcounter{pict@width}
\newcounter{pict@height}
\newlength{\pict@scale}
\newcommand{\psfigadd}[4]{%
\setcounter{pict@width}{1*\ratio{#2+\pict@scale/2}{\pict@scale}}
\setcounter{pict@height}{1*\ratio{#3+\pict@scale/2}{\pict@scale}}
\setlength{\unitlength}{\pict@scale}
\hbox to #2{\hspace{-\fill}\begin{picture}(\thepict@width,\thepict@height)
\put(0,0){\psfig{figure=#1,width=#2,height=#3,clip=}}
\SetScale{0.283466457}
\SetWidth{1.763889}
{#4}
\end{picture}}
}
\newcounter{pict@widthfst}
\newcounter{pict@widthscd}
\newcounter{pict@widthtot}
\newcommand{\psfigaddtwo}[7]{%
\setcounter{pict@widthfst}{1*\ratio{#2+\pict@scale/2}{\pict@scale}}
\setcounter{pict@widthscd}{1*\ratio{#2+#4+\pict@scale/2}{\pict@scale}}
\setcounter{pict@widthtot}{1*\ratio{#2+#4+#6+\pict@scale/2}{\pict@scale}}
\setcounter{pict@height}{1*\ratio{#3+\pict@scale/2}{\pict@scale}}
\setlength{\unitlength}{\pict@scale}
\hbox{\hspace{-\fill}\begin{picture}(\thepict@widthtot,\thepict@height)
\put(0,0){\psfig{figure=#1,width=#2,height=#3,clip=}}
\put(\thepict@widthscd,0){\psfig{figure=#5,width=#6,height=#3,clip=}}
\SetScale{0.283466457}
\SetWidth{1.763889}
{#7}
\end{picture}}
}
\newcommand{\psfigror}[4]{%
\setcounter{pict@width}{1*\ratio{#2+\pict@scale/2}{\pict@scale}}
\setcounter{pict@height}{1*\ratio{#3+\pict@scale/2}{\pict@scale}}
\setlength{\unitlength}{\pict@scale}
\hbox{\begin{picture}(\thepict@width,\thepict@height)
\put(0,\thepict@height){\psfig{figure=#1,width=#3,height=#2,clip=,angle=270}}
\SetScale{0.283466457}
\SetWidth{1.763889}
{#4}
\end{picture}}
}
\newcommand{\psfigrol}[4]{%
\setcounter{pict@width}{1*\ratio{#2+\pict@scale/2}{\pict@scale}}
\setcounter{pict@height}{1*\ratio{#3+\pict@scale/2}{\pict@scale}}
\setlength{\unitlength}{\pict@scale}
\hbox{\begin{picture}(\thepict@width,\thepict@height)
\put(0,0){\psfig{figure=#1,width=#3,height=#2,clip=,angle=90}}
\SetScale{0.283466457}
\SetWidth{1.763889}
{#4}
\end{picture}}
}
\catcode`\@=12 
\newlength\listtextwidth

\catcode`\@=11 
\newlength{\@tabfninsert}
\newlength{\@tabfnwidth}
\newcommand{\tabfootnote}[2]{%
  \setlength{\@tabfninsert}{0.8em}
  \setlength{\@tabfnwidth}{\textwidth}
  \addtolength{\@tabfnwidth}{-\@tabfninsert}
  \addtolength{\@tabfnwidth}{-0.4em}
  \noindent\makebox[\@tabfninsert][r]{\footnotesize$^{#1}$\hfil}\hfill%
  \parbox[t]{\@tabfnwidth}{\footnotesize #2\hfill}}
\catcode`\@=12 

\def\ee{e^+e^-}

\def\be{\begin{equation}}
\def\ee{\end{equation}}
\def\bea{\begin{eqnarray}}
\def\eea{\end{eqnarray}}
\begin{document}
\prepnum{DESY--07--062}

\title{
Multijet production at low $x_{\rm Bj}$\\
in deep inelastic scattering\\
at HERA
}                                                       
                    
\author{ZEUS Collaboration}
\date{11th May, 2007}

\abstract{ 

Inclusive dijet and trijet production in deep inelastic $ep$ scattering has been
measured for $10<Q^2<100$ GeV$^2$ and low Bjorken $x$,
$10^{-4}<x_{\rm Bj}<10^{-2}$.  The data were taken at the HERA $ep$ collider 
with centre-of-mass energy $\sqrt{s} = 318 \gev$ using the ZEUS detector 
and correspond to an integrated luminosity of $82~{\rm pb}^{-1}$.  Jets were 
identified in the hadronic centre-of-mass (HCM) frame using the $k_{T}$ cluster
algorithm in the longitudinally invariant inclusive mode.  Measurements of dijet and trijet 
differential cross sections are presented as functions of $Q^2$, $x_{\rm Bj}$, jet 
transverse energy, and jet pseudorapidity.  As a further examination of low-$x_{\rm Bj}$ dynamics, 
multi-differential cross sections as functions of the jet correlations in transverse momenta, 
azimuthal angles, and pseudorapidity are also presented.  Calculations at $\mathcal{O}(\alpha_{s}^3)$ generally 
describe the trijet data well and improve the description of the dijet data compared to the
calculation at $\mathcal{O}(\alpha_{s}^2)$.

}

\makezeustitle

\def\3{\ss}                                                                                        
\newcommand{\address}{ }                                                                           
\pagenumbering{Roman}                                                                              
                                                   %
\begin{center}                                                                                     
{                      \Large  The ZEUS Collaboration              }                               
\end{center}                                                                                       
  S.~Chekanov$^{   1}$,                                                                            
  M.~Derrick,                                                                                      
  S.~Magill,                                                                                       
  B.~Musgrave,                                                                                     
  D.~Nicholass$^{   2}$,                                                                           
  \mbox{J.~Repond},                                                                                
  R.~Yoshida\\                                                                                     
 {\it Argonne National Laboratory, Argonne, Illinois 60439-4815}, USA~$^{n}$                       
\par \filbreak                                                                                     
  M.C.K.~Mattingly \\                                                                              
 {\it Andrews University, Berrien Springs, Michigan 49104-0380}, USA                               
\par \filbreak                                                                                     
  M.~Jechow, N.~Pavel~$^{\dagger}$, A.G.~Yag\"ues Molina \\                                        
  {\it Institut f\"ur Physik der Humboldt-Universit\"at zu Berlin,                                 
           Berlin, Germany}
% ...
\newpage                                                                                           
$^{\    1}$ supported by DESY, Germany \\                                                          
$^{\    2}$ also affiliated with University College London, UK \\                                  
% ...
$^{\dagger}$ deceased \\                                                                           
%
                                                           %
                                                           %
\begin{tabular}[h]{rp{14cm}}                                                                       
$^{a}$ &  supported by the Natural Sciences and Engineering Research Council of Canada (NSERC) \\  
$^{b}$ &  supported by the German Federal Ministry for Education and Research (BMBF), under        
          contract numbers HZ1GUA 2, HZ1GUB 0, HZ1PDA 5, HZ1VFA 5\\                                
$^{c}$ &  supported in part by the MINERVA Gesellschaft f\"ur Forschung GmbH, the Israel Science   
          Foundation (grant no. 293/02-11.2) and the U.S.-Israel Binational Science Foundation \\  
$^{d}$ &  supported by the German-Israeli Foundation and the Israel Science Foundation\\           
$^{e}$ &  supported by the Italian National Institute for Nuclear Physics (INFN) \\                
$^{f}$ &  supported by the Japanese Ministry of Education, Culture, Sports, Science and Technology 
          (MEXT) and its grants for Scientific Research\\                                          
$^{g}$ &  supported by the Korean Ministry of Education and Korea Science and Engineering          
          Foundation\\                                                                             
$^{h}$ &  supported by the Netherlands Foundation for Research on Matter (FOM)\\                   
$^{i}$ &  supported by the Polish State Committee for Scientific Research, grant no.               
          620/E-77/SPB/DESY/P-03/DZ 117/2003-2005 and grant no. 1P03B07427/2004-2006\\             
$^{j}$ &  partially supported by the German Federal Ministry for Education and Research (BMBF)\\   
$^{k}$ &  supported by RF Presidential grant N 8122.2006.2 for the leading                         
          scientific schools and by the Russian Ministry of Education and Science through its grant
          Research on High Energy Physics\\                                                        
$^{l}$ &  supported by the Spanish Ministry of Education and Science through funds provided by     
          CICYT\\                                                                                  
$^{m}$ &  supported by the Particle Physics and Astronomy Research Council, UK\\                   
$^{n}$ &  supported by the US Department of Energy\\                                               
$^{o}$ &  supported by the US National Science Foundation. Any opinion,                            
findings and conclusions or recommendations expressed in this material                             
are those of the authors and do not necessarily reflect the views of the                           
National Science Foundation.\\                                                                     
$^{p}$ &  supported by the Polish Ministry of Science and Higher Education                         
as a scientific project (2006-2008)\\                                                              
$^{q}$ &  supported by FNRS and its associated funds (IISN and FRIA) and by an Inter-University    
          Attraction Poles Programme subsidised by the Belgian Federal Science Policy Office\\     
$^{r}$ &  supported by the Malaysian Ministry of Science, Technology and                           
Innovation/Akademi Sains Malaysia grant SAGA 66-02-03-0048\\                                       
\end{tabular}                                                                                      
                                                           %
                                                           %

\pagenumbering{arabic} 
\pagestyle{plain}

\section{Introduction}
\label{sec-int}

Multijet production in deep inelastic $ep$ scattering (DIS) at HERA has been 
used to test the predictions of perturbative QCD (pQCD) over a large range of negative four-momentum transfer squared, 
$Q^2$, and to determine the strong coupling constant $\alpha_s$ \cite{epj:c44:183,Adloff:2001kg}.
At leading order (LO) in $\alpha_s$, dijet production in neutral current DIS
proceeds via the boson-gluon-fusion ($V^{*} g \rightarrow q\bar{q}$ with $V=\gamma$, $Z^0$) 
and QCD-Compton ($V^{*} q \rightarrow qg$) processes.
Events with three jets can be seen as dijet processes with 
an additional gluon radiation or with a gluon splitting into a quark-antiquark pair 
and are directly sensitive to $\mathcal{O}(\alpha_s^{2})$ QCD effects.
The higher sensitivity to $\alpha_s$ and the large number of degrees of freedom of the trijet final state 
provide a good testing ground for the pQCD predictions.  In particular, 
multijet production in DIS is an ideal environment for 
investigating different approaches to parton dynamics at low Bjorken-$x$, $x_{\rm Bj}$\cite{pr:179:1547}.  
An understanding of this regime is of particular 
relevance in view of the startup of the LHC, where many of the Standard Model processes such as the production 
of electroweak gauge bosons or the Higgs particle involve the collision of partons with 
a low fraction of the proton momentum.

In the usual collinear QCD factorisation approach, the cross sections are obtained as the convolution of 
perturbative matrix elements and parton densities evolved according to the 
DGLAP evolution equations~\cite{*sovjnp:15:438,*np:b126:298,*sovjnp:20:94,*jetp:46:641}.  These equations 
resum to all orders the terms proportional to $\alpha_s \ln Q^2$ and the double logarithms $\ln Q^2 \cdot \ln{1/x}$, 
where $x$ is the fraction of the proton momentum carried by a parton, which is equal to $x_{\rm Bj}$ in the 
quark-parton model.  In the DGLAP approach, the parton participating in the hard scattering is 
the result of a partonic cascade ordered in transverse momentum, $p_T$.  The partonic cascade starts from a low-$p_T$ 
and high-$x$ parton from the incoming proton and ends up, after consecutive branching, in the high-$p_T$ 
and low-$x$ parton entering in the hard scattering.  
This approximation has been tested extensively at HERA and was found to describe well the inclusive 
cross sections~\cite{epj:c21:33,pr:d67:012007} and jet production
~\cite{epj:c44:183,Adloff:2001kg,epj:c23:13,*pl:b547:164,*epj:c21:443,epj:c19:289,*pl:b515:17,*np:b470:3}.  
At low $x_{\rm Bj}$, where the phase space for parton emissions increases, terms proportional to 
$\alpha_s \ln 1/x$ may become large and spoil the accuracy of the DGLAP approach.  In this region the transverse momenta and 
angular correlations between partons produced in the hard scatter may be sensitive to effects beyond
DGLAP dynamics.  The information about cross sections, transverse energy, $E_T$, and angular correlations between the two 
leading jets in multijet production therefore provides an important testing 
ground for studying the parton dynamics in the region of small $x_{\rm Bj}$. 

In this analysis, correlations for both azimuthal and polar angles, and correlations in jet transverse 
energy and momenta for dijet and trijet production in the hadronic $(\gamma^{*}p)$ centre-of-mass (HCM) frame are measured 
with high statistical precision in the kinematic region restricted to $10 < Q^2 < 100\gev^2$ and 
$10^{-4} < x_{\rm Bj} < 10^{-2}$.  The results are compared with pQCD calculations at next-to-leading order (NLO).  
A similar study of inclusive dijet production was performed by the H1 collaboration~\cite{epj:c37:141}.

\section{Experimental set-up}
\label{sec-det}
The data used in this analysis were collected during the 1998-2000
running period, when HERA operated with protons of energy
$E_p=920$~GeV and electrons or positrons\footnote{In the following,
the term ``electron'' denotes generically both the electron ($e^-$)
and the positron ($e^+$).}  of energy $E_e=27.5$~GeV, and correspond
to an integrated luminosity of $81.7\pm 1.8$~pb$^{-1}$. A detailed
description of the ZEUS detector can be found
elsewhere~\cite{pl:b293:465,zeus:1993:bluebook}. A brief outline of
the components that are most relevant for this analysis is given
below.
 
Charged particles are measured in the central tracking detector
(CTD)~\cite{nim:a279:290,*npps:b32:181,*nim:a338:254}, which operates
in a magnetic field of $1.43\Tesla$ provided by a thin superconducting
solenoid. The CTD consists of 72~cylindrical drift chamber layers,
organised in nine superlayers covering the polar-angle\footnote{The
ZEUS coordinate system is a right-handed Cartesian system, with the
$Z$ axis pointing in the proton beam direction, referred to as the
``forward direction'', and the $X$ axis pointing left towards the
centre of HERA. The coordinate origin is at the nominal interaction
point.}  region \mbox{$15^\circ<\theta<164^\circ$}. The transverse
momentum resolution for full-length tracks can be parameterised as
$\sigma(p_T)/p_T=0.0058p_T\oplus0.0065\oplus0.0014/p_T$, with $p_T$ in
$\gev$. The tracking system was used to measure the interaction vertex
with a typical resolution along (transverse to) the beam direction of
0.4~(0.1)~cm and also to cross-check the energy scale of the
calorimeter.

The high-resolution uranium-scintillator calorimeter
(CAL)~\cite{nim:a309:77,*nim:a309:101,*nim:a321:356,*nim:a336:23}
covers $99.7\%$ of the total solid angle and consists of three parts:
the forward (FCAL), the barrel (BCAL) and the rear (RCAL)
calorimeters. Each part is subdivided transversely into towers and
longitudinally into one electromagnetic section and either one (in
RCAL) or two (in BCAL and FCAL) hadronic sections. The smallest
subdivision of the calorimeter is called a cell. Under test-beam
conditions, the CAL single-particle relative energy resolutions were
$\sigma(E)/E=0.18/\sqrt{E}$ for electrons and
$\sigma(E)/E=0.35/\sqrt{E}$ for hadrons, with $E$ in GeV.
 
The luminosity was measured from the rate of the bremsstrahlung
process $ep\rightarrow e\gamma p$. The resulting small-angle energetic
photons were measured by the luminosity
monitor~\cite{desy-92-066,*zfp:c63:391,*acpp:b32:2025}, a
lead-scintillator calorimeter placed in the HERA tunnel at $Z=-107$ m.

\section{Kinematics and event selection}
\label{sec-sel}

A three-level trigger system was used to select events online 
\cite{zeus:1993:bluebook,proc:chep:1992:222}.
Neutral current DIS events were selected by 
requiring that a scattered electron candidate with an energy more than 4~$\gev$ was measured in
the CAL.  The variable $x_{\rm Bj}$, the inelasticity $y$, and $Q^2$ were reconstructed offline 
using the electron (subscript $e$) \cite{proc:hera:1991:43} and Jacquet-Blondel ({\rm JB}) 
\cite{proc:epfacility:1979:391} methods.  For each event, the reconstruction of the 
hadronic final state was performed using a combination of 
track and CAL information, excluding the cells and the track associated with the 
scattered electron.  The selected tracks and CAL clusters were treated as 
massless energy flow objects (EFOs) \cite{briskin:phd:1998}.  

The offline selection of DIS events was similar to that 
used in the previous ZEUS measurement \cite{epj:c44:183} and was based on the following requirements:

\begin{itemize}
\item $E_e^\prime > 10\gev$, where $E_e^\prime$ is the scattered electron energy after 
correction for energy loss from the inactive material in the detector;

\item $y_e < 0.6$ and $y_{\rm JB} > 0.1$, to ensure 
a kinematic region with good reconstruction;

\item $40 < \delta < 60\gev$, where $\delta = \sum_i(E_i - P_{Z,i})$, where $E_i$ and $P_{Z,i}$ 
are the energy and $z$-momentum of each final-state object.  
The lower cut removed background from photoproduction and events with large initial-state
QED radiation, while the upper cut removed cosmic-ray background;

\item $|Z_{\rm vtx}| < 50$~cm, where $Z_{\rm vtx}$ is the $Z$ position of the
reconstructed primary vertex, to select events consistent with $ep$
collisions.

\end{itemize}

The kinematic range of the analysis is

\begin{center}
$10 < Q^2 < 100 \gev^2$, $10^{-4} < x_{\rm Bj} < 10^{-2}$ and $0.1 < y < 0.6$.
\end{center}

Jets were reconstructed using the $k_T$ cluster algorithm
\cite{np:b406:187} in the longitudinally invariant inclusive mode
\cite{pr:d48:3160}.  The jet search was conducted in the HCM frame, 
which is equivalent to the Breit frame \cite{feynman:1972:photon,*zpf:c2:237} apart from a 
longitudinal boost.

The jet phase space is defined by selection cuts on the jet 
pseudorapidity, $\eta^{\rm jet}_{\rm LAB}$, in the laboratory frame and on the jet
transverse energy, $E^{\rm jet}_{T,\rm HCM}$, in the HCM frame:
\[
-1.0 < \eta^{\rm jet1,2(,3)}_{\rm LAB} < 2.5~{\rm and}~ E^{\rm jet1}_{T,\rm HCM}>7~{\rm GeV}, E^{\rm
jet2(,3)}_{T,\rm HCM}>5~{\rm GeV},
\]

where jet1,2(,3) refers to the two (three) jets with the highest transverse energy 
in the HCM frame for a given event.  The dijet and trijet samples are inclusive in that
they contain at least two or three jets passing the selection criteria, respectively.

\section{Monte Carlo simulation}
\label{mc}
Monte Carlo (MC) simulations were used to correct the data for detector 
effects, inefficiencies of the event selection and the jet reconstruction, as well as for QED effects. 
Neutral current DIS events were generated using the {\sc Ariadne}~4.10 program \cite{cpc:71:15} 
and the {\sc Lepto}~6.5 program \cite{cpc:101:108} interfaced to {\sc Heracles}~4.5.2 \cite{cpc:69:155} 
via {\sc Django}~6.2.4 \cite{cpc:81:381}. The {\sc Heracles} program includes
QED effects up to $\mathcal{O}(\alpha_{\rm EM}^{2})$.
In the case of {\sc Ariadne}, events were generated using the colour-dipole model \cite{np:b306:746}, 
whereas for {\sc Lepto}, the matrix-elements plus parton-shower model was used.  The CTEQ5L parameterisations of the 
proton parton density functions (PDFs)~\cite{pr:d55:1280} 
were used in the generation of DIS events for {\sc Ariadne}, and the CTEQ4D PDFs~\cite{pr:d55:1280} were 
used for {\sc Lepto}.  For hadronisation the Lund string model \cite{prep:97:31}, 
as implemented in {\sc Jetset} 7.4 \cite{cpc:46:43,cpc:82:74} was used. 

The ZEUS detector response was simulated with a program based on {\sc Geant} 3.13
\cite{tech:cern-dd-ee-84-1}. The generated events were passed through the 
detector simulation, subjected to the same trigger requirements as the data,
and processed by the same reconstruction and offline programs.

The measured distributions of the global kinematic variables are well described by both the {\sc Ariadne} 
and {\sc Lepto} MC models after reweighting in $Q^2$\cite{epj:c44:183}.
The {\sc Lepto} simulation gives a better overall description of the 
jet variables, but {\sc Ariadne} provides a better description of dijets with small
azimuthal separation.  Therefore, for this analysis, the events generated with the {\sc Ariadne}
program were used to determine the acceptance corrections. The events generated with
{\sc Lepto} were used to estimate the uncertainty associated with the treatment of
the parton shower.

\section{NLO QCD calculations}
\label{nlo}
The NLO calculations were carried out in the \MSbar scheme for
five massless quark flavors with the program {\sc NLOjet} \cite{prl:87:082001}.
The {\sc NLOjet} program allows a computation of
the dijet (trijet) production cross sections to next-to-leading order, i.e. including all
terms up to $\mathcal{O}(\alpha_s^{2})$ ($\mathcal{O}(\alpha_s^{3})$).  In certain regions of the 
jet phase space, where the two hardest jets are not balanced in transverse momentum, {\sc NLOjet} 
can be used to calculate the cross sections for dijet production at $\mathcal{O}(\alpha_s^{3})$.  
It was checked that the LO and NLO calculations from 
{\sc NLOjet} agree with those of {\sc Disent} \cite{np:b485:291} at the 1-2\% level for the 
dijet cross sections \cite{thesis:krumnack:2004,thesis:li:2004}.

For comparison with the data, the CTEQ6M \cite{Pumplin:2002vw} PDFs were used, 
and the renormalisation and factorisation scales 
were both chosen to be $(\bar{E}_{T,\rm HCM}^2+Q^2)/4$, 
where for dijets (trijets) $\bar{E}_{T,\rm HCM}$ is the average $E_{T,\rm HCM}$ of the 
two (three) highest $E_{T,\rm HCM}$ jets in a given event.
The choice of renormalisation scale matches that used in the previous ZEUS multijet analysis \cite{epj:c44:183}.
The strong coupling constant was set to the value used for the CTEQ6 PDFs,
$\alpha_s(M_Z)=0.118$, and evolved according to the two-loop
solution of the renormalisation group equation. 

The NLO QCD predictions were corrected for hadronisation effects using a
bin-by-bin procedure.  Hadronisation correction factors
were defined for each bin as the ratio of the hadron- to 
parton-level cross sections and were calculated using the {\sc Lepto} MC program, which, at the parton level, 
gives a better agreement with {\sc NLOjet} than {\sc Ariadne}.
The correction factors $C_{\rm had}$ were typically in the range $0.8-0.9$ for most of the phase space.
 
The theoretical uncertainty was estimated by varying the renormalisation scale up and down by a
factor of two.  The uncertainties in the proton PDFs were estimated in the previous ZEUS multijets 
analysis~\cite{epj:c44:183} by repeating {\sc NLOjet} calculations using 40 additional sets from CTEQ6M,
which resulted in a 2.5\% contribution to the theoretical uncertainty and was therefore neglected.

\section{Acceptance Corrections}
\label{sec-unc}
 
The {\sc Ariadne} MC was used to correct the data for detector effects.
The jet transverse energies were corrected for energy losses from inactive material in the detector.  
Typical jet energy correction factors were $1 - 1.2$, 
depending on the transverse energy of the detector-level jet and the jet pseudorapidity.  

The measured cross sections were corrected to the hadron level using a bin-by-bin procedure.  
These corrections account for trigger efficiency, acceptance, and migration.  
Typical efficiencies and purities were about 50\% for the differential cross sections, 
with correction factors typically between 1 and 1.5.  For the double-differential cross sections, 
the efficiencies and purities were typically $20 - 50 \%$, with correction factors between 1 and 2.
 
The cross sections were corrected to the QED Born level by applying
an additional correction obtained from a special sample of the {\sc Lepto} 
MC with the radiative QED effects turned off.  
The QED radiative effects were typically $2 - 4 \%$.

\section{Systematic uncertainties}
\label{sec:sys}
 
A detailed study of the sources contributing to the systematic
uncertainties of the measurements has been performed.  The main
sources contributing to the systematic uncertainties are listed below:

\begin{itemize}
\item the data were corrected using {\sc Lepto} instead of
{\sc Ariadne};
\item the jet energies in the data were scaled up and down by 3\% for jets with transverse 
energy less than $10 \gev$ and 1\% for jets with transverse energy above $10 \gev$, according to
the estimated jet energy scale uncertainty~\cite{epj:c27:531,*Chekanov:2001bw,*Wing:2002fc};
\item the cut on $E_{T,\rm HCM}^{\rm jet}$ for each jet was raised and lowered by 1 GeV, corresponding 
to the $E_T$ resolution;
\item the upper and lower cuts on $\eta^{\rm jet1,2(,3)}_{\rm LAB}$ were each changed by $\pm0.1$,
corresponding to the $\eta$ resolution;
\item the uncertainties due to the selection cuts was estimated by varying the cuts within the 
resolution of each variable.
 
\end{itemize}
 
The largest systematic uncertainties came from the uncertainty of the jet energy
scale, which produced a systematic uncertainty of $5 - 10 \%$.  
For the trijet sample, altering the cut on $E_{T,\rm HCM}^{\rm et3}$ also produced a systematic
uncertainty of $5 - 10 \%$.  The other significant systematic uncertainty arose from the choice of {\sc Lepto}
instead of {\sc Ariadne} for correcting detector effects.  This systematic uncertainty
was also typically $5 - 10 \%$.  The other systematic uncertainties were smaller than or similar to the
statistical uncertainties.
 
The systematic uncertainties not associated with the absolute energy scale of the jets were added 
in quadrature to the statistical uncertainties and are shown as error bars in the figures. The 
uncertainty due to the absolute energy scale of the jets is shown separately as a shaded band in 
each figure, due to the large bin-to-bin correlation.  In addition, there is an overall normalisation 
uncertainty of 2.2\% from the luminosity determination, which is not included in the figures.

\section{Results}
\label{sec-res}

\subsection{Single-differential cross sections 
$d\sigma/dQ^2$, $d\sigma/dx_{\rm Bj}$ and trijet to dijet cross section ratios}

The single-differential cross-sections $d\sigma/dQ^2$ and $d\sigma/dx_{\rm Bj}$
for dijet and trijet production are presented in Figs.~\ref{fig_x_q2}(a) and (c), and 
Tables~\ref{q2_inc_dijet_tab} --~\ref{x_inc_trijet_tab}.  
The ratio $\sigma_{\rm trijet}/\sigma_{\rm dijet}$ of the trijet cross section to the 
dijet cross section, as a function of $Q^2$ and of $x_{\rm Bj}$ are presented in Figs.~\ref{fig_x_q2}(b) and 
~\ref{fig_x_q2}(d), respectively.  The ratio $\sigma_{\rm trijet}/\sigma_{\rm dijet}$ 
is almost $Q^2$ independent, as shown in Fig.~\ref{fig_x_q2}(b), 
and falls steeply with increasing $x_{\rm Bj}$, as shown in Fig.~\ref{fig_x_q2}(d).  
In the cross-section ratios, the experimental and theoretical uncertainties partially cancel, 
providing a possibility to test the pQCD calculations more precisely than can be done with the 
individual cross sections.  Both the cross sections and the cross-section ratios are well described 
by the {\sc NLOjet} calculations.

\subsection{Transverse energy and pseudorapidity dependencies of cross sections}

The single-differential cross-sections $d\sigma/dE_{T,\rm HCM}^{\rm jet}$ for two (three) 
jet events are presented in Fig.~\ref{fig_et}.  The measured cross sections are well described 
by the {\sc NLOjet} calculations over the whole range in $E_{T,\rm HCM}^{\rm jet}$ considered.

The single-differential cross sections $d\sigma/d\eta_{{\rm LAB}}^{\rm jet}$ for dijet and trijet 
production are presented in Figs.~\ref{fig_eta}(a) and~\ref{fig_eta}(c).  
For this figure, the two (three) jets with highest $E_{T,\rm HCM}^{\rm jet}$ were ordered in $\eta_{{\rm LAB}}^{\rm jet}$.  
Also shown are the measurements of the single-differential cross-sections 
$d\sigma/d|\Delta\eta_{\rm HCM}^{\rm jet 1,2}|$, where $|\Delta\eta_{\rm HCM}^{\rm jet 1,2}|$ is the 
absolute difference in pseudorapidity of the two jets with highest $E_{T,\rm HCM}^{\rm jet}$ 
(see Figs.~\ref{fig_eta}(b) and~\ref{fig_eta}(d)).  The {\sc NLOjet} predictions describe the measurements well.

\subsection{Jet transverse energy and momentum correlations}

Correlations in transverse energy of the jets have been investigated by measuring the 
double-differential cross-sections $d^2\sigma/dx_{\rm Bj} d\Delta E_{T,\rm HCM}^{\rm jet 1,2}$, 
where $\Delta E_{T,\rm HCM}^{\rm jet 1,2}$ is the difference in transverse energy between the two jets with the
highest $E_{T,\rm HCM}^{\rm jet}$.  The measurement was performed in $x_{\rm Bj}$ bins, 
which are defined in Table~\ref{x_inc_dijet_tab}, for dijet and trijet production.
Figures~\ref{fig_delet_dijet} and~\ref{fig_delet_trijet} show the cross-sections 
$d^2\sigma/dx_{\rm Bj} d\Delta E_{T,\rm HCM}^{\rm jet 1,2}$ for all bins in $x_{\rm Bj}$ for the dijet and 
trijet samples, respectively.

The {\sc NLOjet} calculations at $\mathcal{O}(\alpha_s^{2})$ do not describe the high-$\Delta E_{T,\rm HCM}^{\rm jet 1,2}$ 
tail of the dijet sample at low $x_{\rm Bj}$, where the calculations fall below the data.  Since these 
calculations give the lowest-order non-trivial contribution to the cross section in 
the region $\Delta E_{T,\rm HCM}^{\rm jet 1,2}>0$, they are affected by large uncertainties 
from the higher-order terms in $\alpha_s$.  A higher-order calculation 
for the dijet sample is possible with {\sc NLOjet} if the region $\Delta E_{T,\rm HCM}^{\rm jet 1,2}$ near zero is avoided.  
{\sc NLOjet} calculations at $\mathcal{O}(\alpha_s^{3})$ for 
the dijet sample have been obtained for the region $\Delta E_{T,\rm HCM}^{\rm jet 1,2}>4\gev$ and are compared to 
the data in Fig.~\ref{fig_delet_dijet}.  With the inclusion of the next 
term in the perturbative series in $\alpha_s$, the {\sc NLOjet} calculations describe the data within the 
theoretical uncertainties.  The {\sc NLOjet} calculations at $\mathcal{O}(\alpha_s^{3})$ for trijet production 
are consistent with the measurements.

As a refinement to the studies of the correlations between the transverse energies of the jets, further correlations 
of the jet tranverse momenta have been investigated.  The correlations in jet transverse momenta were examined by 
measuring two sets of double-differential cross sections:  
$d^2\sigma/dx_{\rm Bj} d|\Sigma \vec{p}_{T,\rm HCM}^{~\rm jet 1,2}|$ and 
$d^2\sigma/dx_{\rm Bj} d(|\Delta \vec{p}_{T,\rm HCM}^{~\rm jet 1,2}|/(2 E_{T,\rm HCM}^{\rm jet 1}))$. The variable 
$|\Sigma \vec{p}_{T,\rm HCM}^{~\rm jet 1,2}|$ is the transverse component of the vector sum of the jet 
momenta of the two jets with the highest $E_{T,\rm HCM}^{\rm jet}$.  
For events with only two jets $|\Sigma \vec{p}_{T,\rm HCM}^{~\rm jet 1,2}| = 0$, and 
additional QCD radiation increases this value.  The variable 
$|\Delta \vec{p}_{T,\rm HCM}^{~\rm jet 1,2}|/(2 E_{T,\rm HCM}^{\rm jet 1})$
is the magnitude of the vector difference of the transverse momenta of the two jets with 
the highest $E_{T,\rm HCM}^{\rm jet}$ scaled by twice the transverse energy of the hardest jet.  
For events with only two jets $|\Delta \vec{p}_{T,\rm HCM}^{~\rm jet 1,2}|/(2 E_{T,\rm HCM}^{\rm jet 1}) = 1$, and additional QCD radiation decreases this value.  
Figures~\ref{fig_sumpt_dijet} --~\ref{fig_deletn_trijet} show the cross-sections 
$d^2\sigma/dx_{\rm Bj} d|\Sigma \vec{p}_{T,\rm HCM}^{~\rm jet 1,2}|$ and the cross-sections 
$d^2\sigma/dx_{\rm Bj} d|\Delta \vec{p}_{T,\rm HCM}^{~\rm jet 1,2}|/(2 E_{T,\rm HCM}^{\rm jet 1})$ 
in bins of $x_{\rm Bj}$ for the dijet and trijet samples.

At low $x_{\rm Bj}$, the {\sc NLOjet} calculations at $\mathcal{O}(\alpha_s^{2})$ underestimate the dijet 
cross sections at high values of $|\Sigma \vec{p}_{T,\rm HCM}^{~\rm jet 1,2}|$ and low values of 
$|\Delta \vec{p}_{T,\rm HCM}^{~\rm jet 1,2}|/(2 E_{T,\rm HCM}^{\rm jet 1})$.  
The description of the data by the {\sc NLOjet} calculations at 
$\mathcal{O}(\alpha_s^{2})$ improves at higher values of $x_{\rm Bj}$.  
A higher-order calculation with {\sc NLOjet} at 
$\mathcal{O}(\alpha_s^{3})$ for the dijet sample has been obtained for the region 
$|\Sigma \vec{p}_{T,\rm HCM}^{~\rm jet 1,2}|>4\gev$, which is compared to the data in 
Fig.~\ref{fig_sumpt_dijet}; and for the region 
$|\Delta \vec{p}_{T,\rm HCM}^{~\rm jet 1,2}|/(2 E_{T,\rm HCM}^{\rm jet 1})<0.85$, 
which is compared to the data in Fig.~\ref{fig_deletn_dijet}.
With the inclusion of the next term in the perturbative series in 
$\alpha_s$, the {\sc NLOjet} calculations describe the data well.  The {\sc NLOjet} calculations 
at $\mathcal{O}(\alpha_s^{3})$ for trijet production are consistent with the measurements.

\subsection{Azimuthal distributions of the jets}

Measurements of the double-differential cross-section $d^2\sigma/dx_{\rm Bj}d|\Delta\phi_{\rm HCM}^{\rm jet 1,2}|$, 
where $|\Delta\phi_{\rm HCM}^{\rm jet 1,2}|$ is the azimuthal separation of 
the two jets with the largest $E_{T,\rm HCM}^{\rm jet}$, for dijet and trijet production are shown in 
Figs.~\ref{fig_delphi_dijet} and~\ref{fig_delphi_trijet} for all bins in $x_{\rm Bj}$.  
For both dijet and trijet production the cross section falls with $|\Delta\phi_{\rm HCM}^{\rm jet 1,2}|$.  
The {\sc NLOjet} calculations at $\mathcal{O}(\alpha_S^{2})$ for dijet production decrease more rapidly with 
$|\Delta\phi_{\rm HCM}^{\rm jet 1,2}|$ than the data and the calculations disagree with the data at 
low $|\Delta\phi_{\rm HCM}^{\rm jet 1,2}|$.  A higher-order {\sc NLOjet} calculation at $\mathcal{O}(\alpha_S^{3})$ 
for the dijet sample has been obtained for the region $|\Delta\phi_{\rm HCM}^{\rm jet 1,2}| < 3\pi / 4$ and describes the 
data well.  The measurements for trijet production are reasonably well described by the {\sc NLOjet} calculations at 
$\mathcal{O}(\alpha_S^{3})$.

A further investigation has been performed by measuring the cross-section $d^2\sigma/dQ^2dx_{\rm Bj}$ for dijet 
(trijet) events with $|\Delta\phi_{\rm HCM}^{\rm jet 1,2}|< 2\pi / 3$ as a function of $x_{\rm Bj}$.  
For the two-jet final states, the presence of two leading jets with $|\Delta\phi_{\rm HCM}^{\rm jet 1,2}| < 2\pi / 3$ 
can indicate another high-$E_T$ jet or set of high-$E_T$ jets outside the measured $\eta$ range. 
These cross sections are presented in Fig.~\ref{fig_low_delphi}. 
The {\sc NLOjet} calculations at $\mathcal{O}(\alpha_S^{2})$ for dijet production underestimate the data, 
the difference increasing towards low $x_{\rm Bj}$.  The {\sc NLOjet} calculations at $\mathcal{O}(\alpha_S^{3})$ 
are up to about one order of magnitude larger than the $\mathcal{O}(\alpha_S^{2})$ calculations and are consistent 
with the data, demonstrating the importance of the higher-order terms in the description of the 
data especially at low $x_{\rm Bj}$.  The {\sc NLOjet} calculations at $\mathcal{O}(\alpha_S^{3})$ 
describe the trijet data within the renormalisation-scale uncertainties.

\section{Summary}

Dijet and trijet production in deep inelastic $ep$ scattering has been
measured in the phase space region $10<Q^2<100$ GeV$^2$ and $10^{-4} <x_{\rm Bj}< 10^{-2}$ 
using an integrated luminosity of 82~pb$^{-1}$ collected by the ZEUS experiment.  
The high statistics have made possible detailed studies 
of multijet production at low $x_{\rm Bj}$. The dependence of dijet and 
trijet production on the kinematic variables $Q^2$ and $x_{\rm Bj}$ and on 
the jet variables $E_{T,\rm HCM}^{\rm jet}$ and $\eta_{{\rm LAB}}^{\rm jet}$ is well described by 
perturbative QCD calculations which include NLO corrections.  
To investigate possible deviations with respect to 
the collinear factorisation approximation used in the standard 
pQCD approach, measurements of the correlations between the two 
jets with highest $E_{T,\rm HCM}^{\rm jet}$ have been made.  At low $x_{\rm Bj}$, 
measurements of dijet production with low azimuthal separation are reproduced by 
the perturbative QCD calculations provided that higher-order 
terms ($\mathcal{O}(\alpha_s^{3})$) are accounted for.  Such terms increase the predictions of 
pQCD calculations by up to one order of magnitude when the two jets with the highest 
$E_{T,\rm HCM}^{\rm jet 1,2}$ are not balanced in transverse momentum.  This demonstrates the 
importance of higher-order corrections in the low-$x_{\rm Bj}$ region.

\section*{Acknowledgements}
It is a pleasure to thank the DESY Directorate for their strong
support and encouragement. The remarkable achievements of the HERA
machine group were essential for the successful completion of this
work and are greatly appreciated. The design, construction and
installation of the ZEUS detector has been made possible by the
efforts of many people who are not listed as authors.  It is also a pleasure
to thank Zoltan Nagy for useful discussions about {\sc NLOjet}.

\nocite{thesis:gosau:2007, thesis:danielson:2007}

{
\def\bibname{\Large\bf References}
\def\refname{\Large\bf References}
\pagestyle{plain}
\ifzeusbst
  \bibliographystyle{../BiBTeX/bst/l4z_default}
\fi
\ifzdrftbst
  \bibliographystyle{../BiBTeX/bst/l4z_draft}
\fi
\ifzbstepj
  \bibliographystyle{../BiBTeX/bst/l4z_epj}
\fi
\ifzbstnp
  \bibliographystyle{../BiBTeX/bst/l4z_np}
\fi
\ifzbstpl
  \bibliographystyle{../BiBTeX/bst/l4z_pl}
\fi
{\raggedright
\bibliography{../BiBTeX/user/syn.bib,%
              ../BiBTeX/user/ADAM.bib,%
              ../BiBTeX/bib/l4z_articles.bib,%
              ../BiBTeX/bib/l4z_books.bib,%
              ../BiBTeX/bib/l4z_conferences.bib,%
              ../BiBTeX/bib/l4z_h1.bib,%
              ../BiBTeX/bib/l4z_misc.bib,%
              ../BiBTeX/bib/l4z_old.bib,%
              ../BiBTeX/bib/l4z_preprints.bib,%
              ../BiBTeX/bib/l4z_replaced.bib,%
              ../BiBTeX/bib/l4z_temporary.bib,%
              ../BiBTeX/bib/l4z_zeus.bib}
}
\vfill\eject

\begin{table}[htbp]
  \begin{center}
    \begin{tabular}{||c|cccc||c||c||}
      \hline 
$Q^2$ & $\frac{d\sigma}{dQ^2}$ & $\delta_{stat}$ & $\delta_{syst}$ & $\delta_{ES}$ & $C_{QED}$ & $C_{had}$ \\ 
$(\gev^2)$ & $({\rm pb}/\gev^2)$ & $({\rm pb}/\gev^2)$ & $({\rm pb}/\gev^2)$ & $({\rm pb}/\gev^2)$ & & \\ 
      \hline 
      \hline 
10 - 15&$66.0$&$0.8$&$ ^{+3.7}_{-4.4}$&$ ^{+5.7}_{-5.9}$&$0.984$&$0.866$\\ 
15 - 20&$41.4$&$0.6$&$ ^{+2.0}_{-2.4}$&$ ^{+3.5}_{-3.6}$&$0.968$&$0.870$\\ 
20 - 30&$26.2$&$0.3$&$ ^{+1.0}_{-0.8}$&$ ^{+2.2}_{-2.0}$&$0.965$&$0.876$\\ 
30 - 50&$14.0$&$0.1$&$ ^{+0.4}_{-0.3}$&$ ^{+1.0}_{-1.1}$&$0.955$&$0.884$\\ 
50 - 100&$5.82$&$0.06$&$ ^{+0.17}_{-0.16}$&$ ^{+0.38}_{-0.38}$&$0.952$&$0.887$\\ 
      \hline 
    \end{tabular}
    \caption{The inclusive dijet cross sections as functions of $Q^{2}$.  Included are the statistical, systematic, and jet energy scale uncertainties in columns 3, 4, and 5, respectively.  Column 6 shows the correction factor from QED radiative effects applied to the measured cross sections, and column 7 shows the hadronization correction applied to the {\sc NLOjet} calculations shown in the figures.}
    \label{q2_inc_dijet_tab}
  \end{center}
\end{table}
\begin{table}[htbp]
  \begin{center}
    \begin{tabular}{||c|cccc||c||c||}
      \hline 
$x_{\rm Bj}\times 10^{-4}$ & $\frac{d\sigma}{dx_{\rm Bj}}$ & $\delta_{stat} $ & $\delta_{syst}$ & $\delta_{ES}$ & $C_{QED}$ & $C_{had}$ \\ 
& $({\rm pb},\times 10^{-4})$ & $({\rm pb},\times 10^{-4})$ & $({\rm pb},\times 10^{-4})$ & $({\rm pb},\times 10^{-4})$ & & \\ 
      \hline 
      \hline 
1.7 - 3.0&$85.3$&$1.7$&$ ^{+5.6}_{-6.8}$&$ ^{+7.0}_{-6.3}$&$0.987$&$0.910$\\ 
3.0 - 5.0&$113.8$&$1.5$&$ ^{+5.9}_{-6.2}$&$ ^{+8.8}_{-8.9}$&$0.975$&$0.887$\\ 
5.0 - 10.0&$83.1$&$0.8$&$ ^{+3.3}_{-3.7}$&$ ^{+6.9}_{-7.1}$&$0.969$&$0.876$\\ 
10.0 - 25.0&$29.5$&$0.3$&$ ^{+0.8}_{-0.8}$&$ ^{+2.2}_{-2.2}$&$0.958$&$0.876$\\ 
25.0 - 100.0&$2.31$&$0.03$&$ ^{+0.08}_{-0.07}$&$ ^{+0.17}_{-0.17}$&$0.948$&$0.862$\\ 
      \hline 
    \end{tabular}
    \caption{The inclusive dijet cross sections as functions of $x_{\rm Bj}$.  Other details as in the caption to Table 1.}
    \label{x_inc_dijet_tab}
  \end{center}
\end{table}
\clearpage
\begin{table}[htbp]
  \begin{center}
    \begin{tabular}{||c|cccc||c||c||}
      \hline 
$Q^2$ & $\frac{d\sigma}{dQ^2}$ & $\delta_{stat}$ & $\delta_{syst}$ & $\delta_{ES}$ & $C_{QED}$ & $C_{had}$ \\ 
$(\gev^2)$ & $({\rm pb}/\gev^2)$ & $({\rm pb}/\gev^2)$ & $({\rm pb}/\gev^2)$ & $({\rm pb}/\gev^2)$ & & \\ 
      \hline 
      \hline 
10 - 15&$7.9$&$0.2$&$ ^{+1.1}_{-1.3}$&$ ^{+1.0}_{-1.0}$&$0.991$&$0.759$\\ 
15 - 20&$4.40$&$0.17$&$ ^{+0.46}_{-0.66}$&$ ^{+0.45}_{-0.52}$&$0.946$&$0.776$\\ 
20 - 30&$3.19$&$0.11$&$ ^{+0.27}_{-0.37}$&$ ^{+0.38}_{-0.38}$&$0.969$&$0.786$\\ 
30 - 50&$1.68$&$0.06$&$ ^{+0.13}_{-0.11}$&$ ^{+0.20}_{-0.19}$&$0.949$&$0.794$\\ 
50 - 100&$0.719$&$0.024$&$ ^{+0.044}_{-0.027}$&$ ^{+0.077}_{-0.070}$&$0.956$&$0.795$\\ 
      \hline 
    \end{tabular}
    \caption{The inclusive trijet cross sections as functions of $Q^{2}$.  Other details as in the caption to Table 1.}
    \label{q2_inc_trijet_tab}
  \end{center}
\end{table}
\begin{table}[htbp]
  \begin{center}
    \begin{tabular}{||c|cccc||c||c||}
      \hline 
$x_{\rm Bj}\times 10^{-4}$ & $\frac{d\sigma}{dx_{\rm Bj}}$ & $\delta_{stat} $ & $\delta_{syst}$ & $\delta_{ES}$ & $C_{QED}$ & $C_{had}$ \\ 
& $({\rm pb},\times 10^{-4})$ & $({\rm pb},\times 10^{-4})$ & $({\rm pb},\times 10^{-4})$ & $({\rm pb},\times 10^{-4})$ & & \\ 
      \hline 
      \hline 
1.7 - 3.0&$14.7$&$0.7$&$ ^{+1.5}_{-3.3}$&$ ^{+1.5}_{-1.9}$&$1.00$&$0.811$\\ 
3.0 - 5.0&$15.9$&$0.5$&$ ^{+2.0}_{-2.3}$&$ ^{+1.9}_{-1.8}$&$0.968$&$0.796$\\ 
5.0 - 10.0&$9.6$&$0.3$&$ ^{+0.9}_{-0.9}$&$ ^{+1.1}_{-1.1}$&$0.961$&$0.780$\\ 
10.0 - 25.0&$3.35$&$0.10$&$ ^{+0.21}_{-0.19}$&$ ^{+0.40}_{-0.37}$&$0.954$&$0.785$\\ 
25.0 - 100.0&$0.192$&$0.013$&$ ^{+0.032}_{-0.020}$&$ ^{+0.023}_{-0.022}$&$0.95$&$0.739$\\ 
      \hline 
    \end{tabular}
    \caption{The inclusive trijet cross sections as functions of $x_{\rm Bj}$.  Other details as in the caption to Table 1.}
    \label{x_inc_trijet_tab}
  \end{center}
\end{table}
\clearpage
\clearpage
\begin{table}[htbp]
  \begin{center}
    \begin{tabular}{||c|c|c||}
      \hline
      Variable & Bin & Boundaries \\
      \hline
      \hline
      $\Delta E_{T,\rm HCM}^{\rm jet 1,2}$ & $1$ & $ 0 - 4~\gev $ \\
      & $2$ & $4 - 10~\gev $ \\
      & $3$ & $10 - 18~\gev $ \\
      & $4$ & $18 - 100~\gev $ \\
      \hline
      $|\Sigma \vec{p}_{T,\rm HCM}^{~\rm jet 1,2}|$ & $1$ & $ 0 - 4~\gev $ \\
      & $2$ & $4 - 10~\gev $ \\
      & $3$ & $10 - 16~\gev $ \\
      & $4$ & $16 - 100~\gev $ \\
      \hline
      $|\Delta \vec{p}_{T,\rm HCM}^{~\rm jet 1,2}|/2 E_{T,\rm HCM}^{\rm jet 1}$ & $1$ & $0 - 0.5$ \\
      & $2$ & $0.5 - 0.7$ \\
      & $3$ & $0.7 - 0.85$ \\
      & $4$ & $0.85 - 1$ \\
      \hline
      $|\Delta \phi_{\rm HCM}^{\rm jet 1,2}|$ & $1$ & $0 - \pi / 4$ \\
      & $2$ & $\pi / 4 - \pi / 2$ \\
      & $3$ & $\pi / 2 - 3 \pi / 4$ \\
      & $4$ & $3 \pi / 4 - \pi$ \\
      \hline
      \hline
    \end{tabular}
    \caption{The bin edges used for the measurements of the jet correlations presented.  For the trijet
      sample, the first two bins in $|\Delta \phi_{\rm HCM}^{\rm jet 1,2}|$ are combined.}
    \label{bin_borders_tab}
  \end{center}
\end{table}
\clearpage


                                %
\begin{figure}[htp]
  \begin {center}
    {\fontfamily{phv}\selectfont \Huge\textbf{ZEUS}}
  \end {center}
  
  \begin{center}                                %
    \begin{minipage}{\linewidth}
      \psfigadd{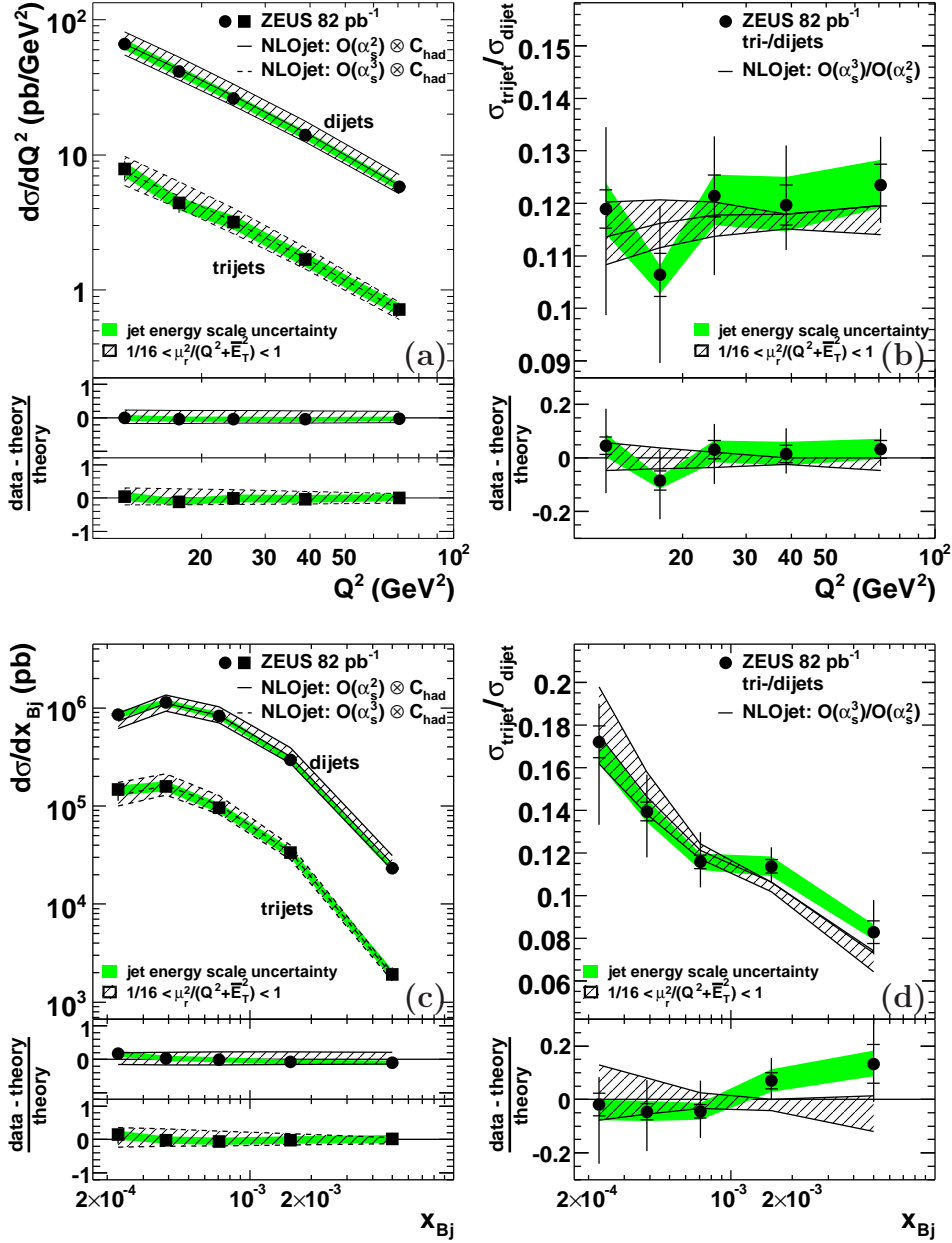}{.8\textwidth}{1.0667\textwidth}{%
        \Text(570,1180)[]{\bf (a)} \Text(1205,1180)[]{\bf(b)} \Text(570,325)[]{\bf (c)} \Text(1205,325)[]{\bf (d)}}
    \end{minipage}
  \end{center}
  \caption{
    Inclusive dijet and trijet cross sections as functions of (a) $Q^{2}$ and (c) $x_{Bj}$.  
    Figures (b) and (d) show the ratios of the trijet to dijet cross sections.  The bin-averaged 
    differential cross sections are plotted at the bin
    centers.  The inner error bars represent the statistical uncertainties. The outer error bars
    represent the quadratic sum of statistical and systematic uncertainties
    not associated with the jet energy scale. The shaded band
    indicates the jet energy scale uncertainty. The predictions
    of perturbative QCD at NLO, corrected for hadronisation
    effects and using the CTEQ6 parameterisations of the proton PDFs, are compared to data.  
    The lower parts of the plots show the relative difference between the data and the corresponding 
    theoretical prediction.  
    The hatched band represents the renormalisation-scale uncertainty of the QCD calculation.
    }
  \label{fig_x_q2}
\end{figure}
                                %
\begin{figure}[htbp]
  \begin {center}
    {\fontfamily{phv}\selectfont \Huge\textbf{ZEUS}}
  \end {center}
  \begin{center}
    \begin{minipage}{\linewidth}
      \psfigadd{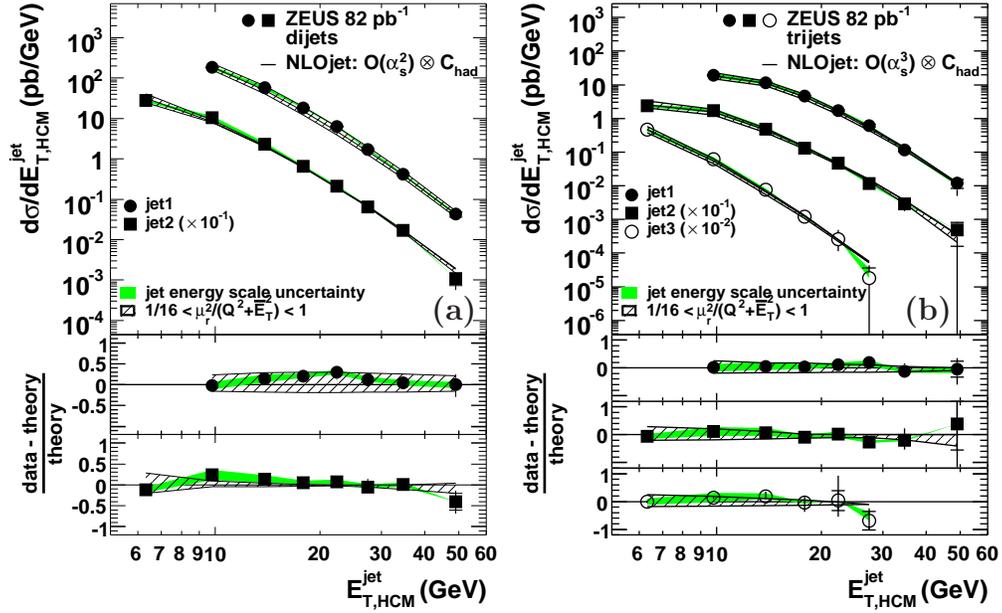}{0.8333\textwidth}{0.5556\textwidth}{%
        \Text(590,430)[]{\bf (a)} \Text(1220,430)[]{\bf (b)}}
    \end{minipage}
  \end{center}
  
  \caption{
    Inclusive dijet (a) and trijet (b)  cross sections as functions of $E_{T,{\rm HCM}}^{\rm jet}$ with the jets 
    ordered in $E_{T,{\rm HCM}}^{\rm jet}$.  The cross sections of the
    second and third jet were scaled for readability.
    Other details as in the caption to Fig. 1.
    }
  \label{fig_et}
\end{figure}
                                %
                                %
\begin{figure}[htbp]
  \begin {center}
    {\fontfamily{phv}\selectfont \Huge\textbf{ZEUS}}
  \end {center}
  
  \begin{center}
    \begin{minipage}{\linewidth}
      \psfigadd{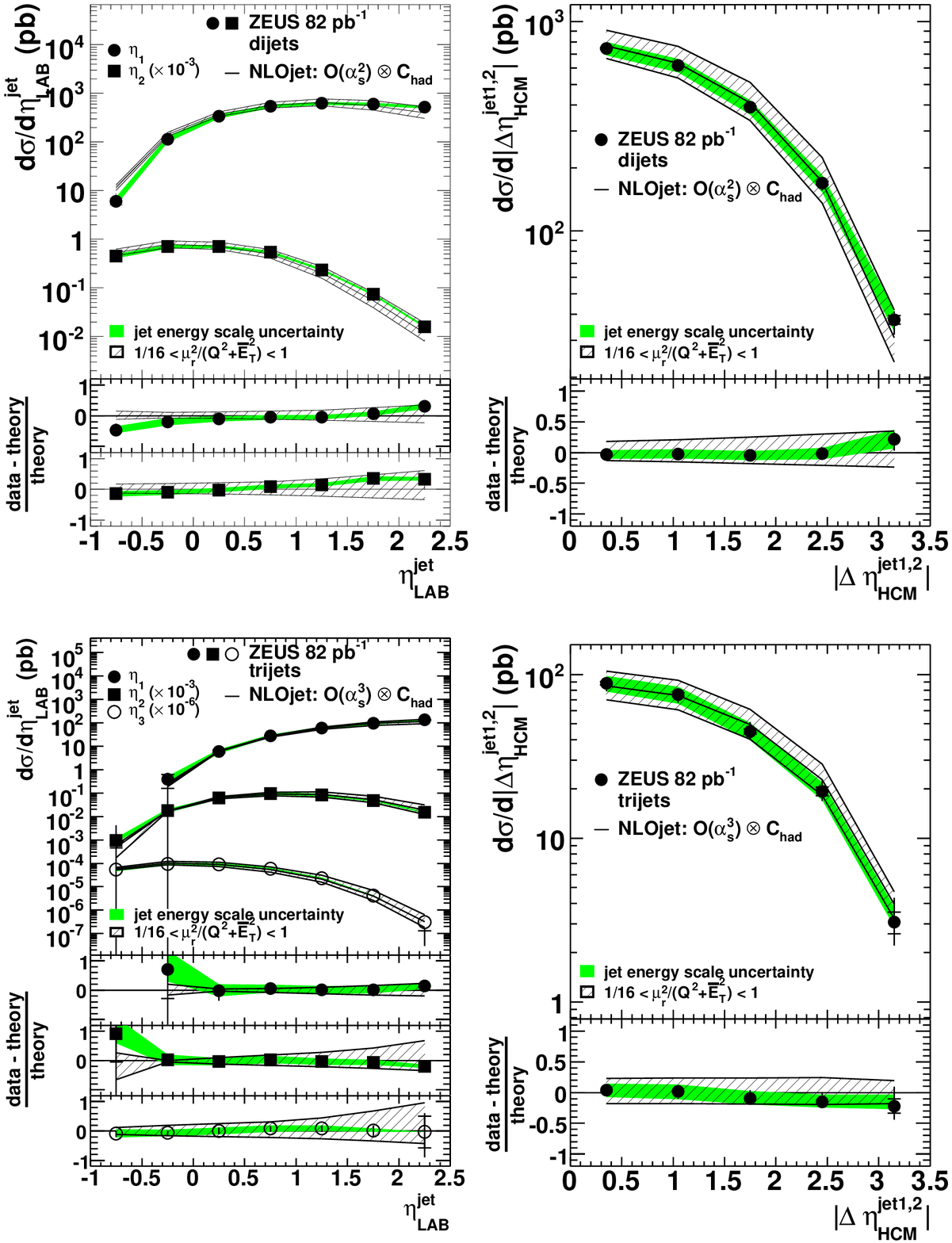}{0.8333\textwidth}{1.1111\textwidth}{%
        \Text(590,1230)[]{\bf (a)} \Text(1200,1230)[]{\bf(b)} \Text(550,430)[]{\bf(c)} \Text(1260,340)[]{\bf(d)}}
    \end{minipage}
  \end{center}
  \caption{ 
    The inclusive dijet (a) and trijet (c) cross sections as
    functions of $\eta^{\rm jet}_{{\rm LAB}}$ with the jets ordered in $\eta^{\rm jet}_{{\rm LAB}}$: 
    $\eta_{{\rm LAB}}^{\rm jet 1} > \eta_{{\rm LAB}}^{\rm jet 2} > \eta_{{\rm LAB}}^{\rm jet 3}$. 
    The cross sections of the second and third jet were scaled for readability.
    Figures (b) and (d) show the dijet and trijet cross sections as
    functions of $|\Delta\eta_{{\rm HCM}}^{\rm jet 1,2}|$ between the two jets with highest $E_{T,\rm HCM}^{\rm jet}$.
    Other details as in the caption in Fig. 1.
    }
  \label{fig_eta}
\end{figure}
                                %
                                %
                                %
\begin{figure}[htbp]
  
  \begin {center}
    {\fontfamily{phv}\selectfont \Huge\textbf{ZEUS}}
  \end {center}
  
  \begin{center}
    \begin{minipage}{\linewidth}
      \psfigadd{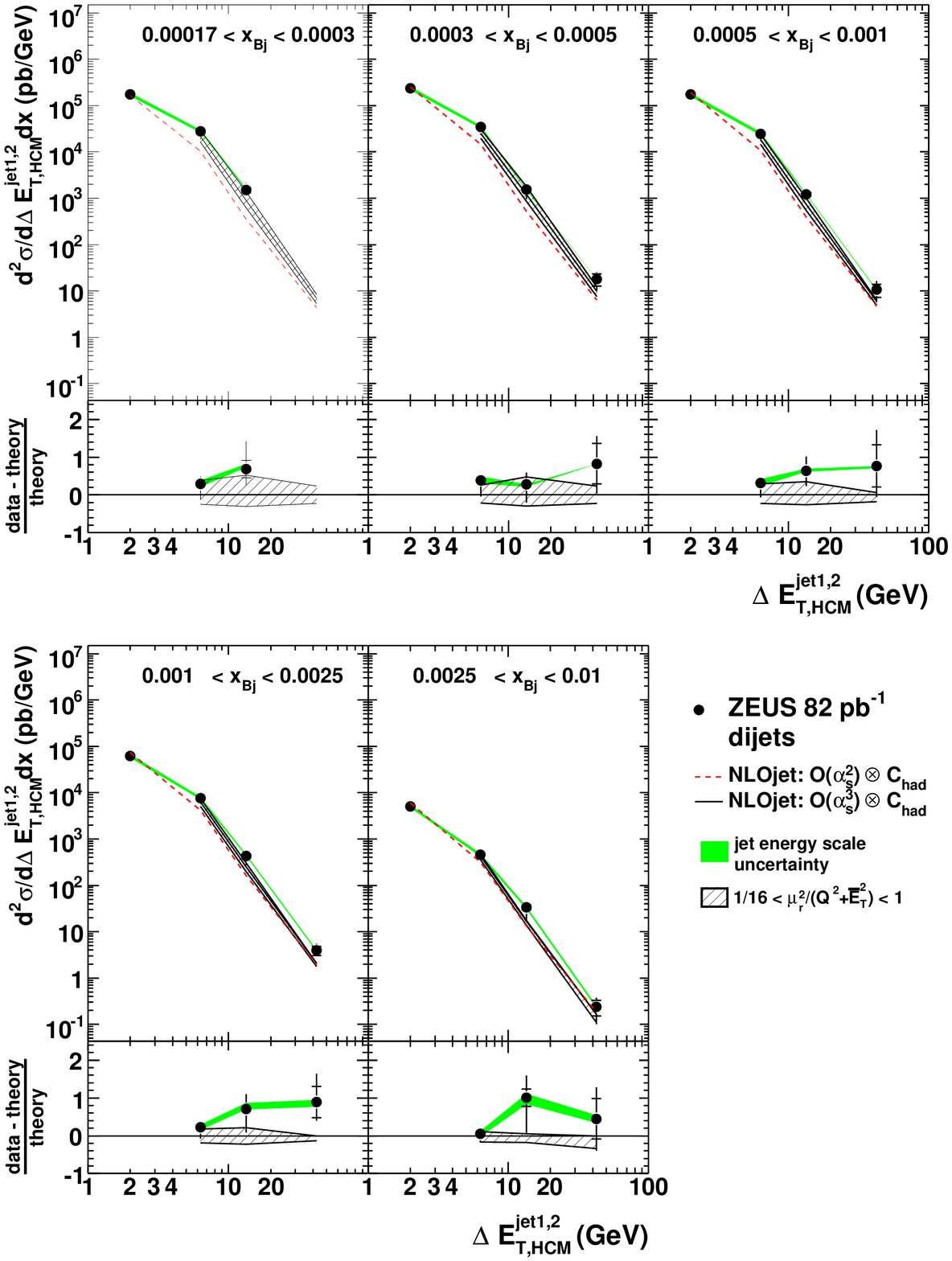}{0.9\textwidth}{1.2\textwidth}{}
    \end{minipage}
  \end{center}
  \caption{ Dijet cross sections as functions of
    $\Delta E_{T,{\rm HCM}}^{\rm jet 1,2}$.  
    The {\sc NLOjet}
    calculations at $\mathcal{O}(\alpha_s^{2})$ $(\mathcal{O}(\alpha_s^{3}))$ are shown
    as dashed (solid) lines.  
    The lower parts of the plots show the relative difference between the data and the $\mathcal{O}(\alpha_s^{3})$ predictions.  
    The boundaries for the bins in $\Delta E_{T,{\rm HCM}}^{\rm jet 1,2}$ are given in Table~\ref{bin_borders_tab}.
    Other details as in the caption to Fig. 1.
    }
  \label{fig_delet_dijet}
\end{figure}
                                %
                                %
                                %
\begin{figure}[htbp]

  \begin {center}
    {\fontfamily{phv}\selectfont \Huge\textbf{ZEUS}}
  \end {center}

  \begin{center}
    \begin{minipage}{\linewidth}
      \psfigadd{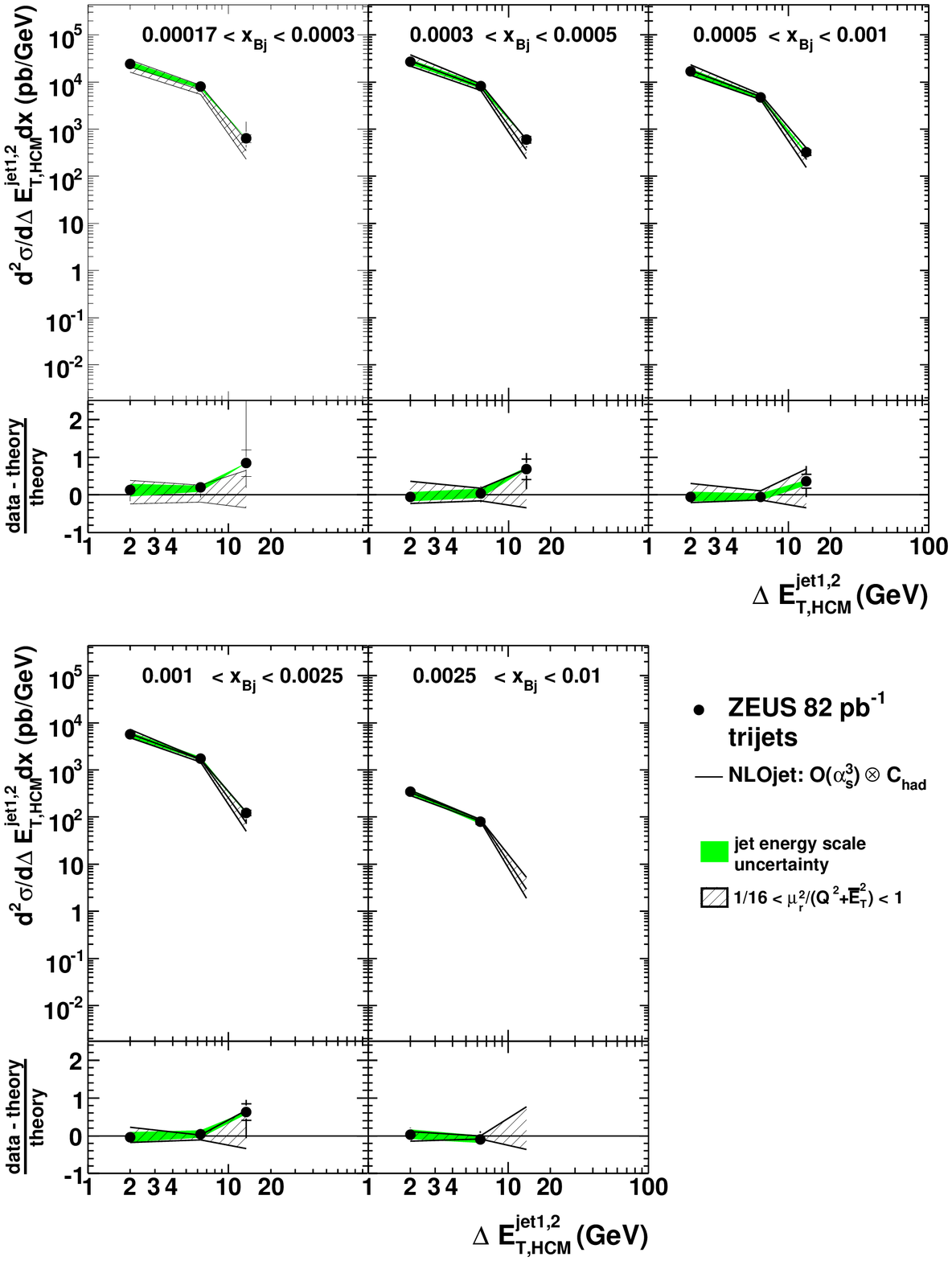}{0.9\textwidth}{1.2\textwidth}{}
    \end{minipage}
  \end{center}
  \caption{ Trijet cross sections as functions of
    $\Delta E_{T,{\rm HCM}}^{\rm jet 1,2}$.  
    The measurements are compared to {\sc NLOjet}
    calculations at $\mathcal{O}(\alpha_s^{3})$. 
    The boundaries for the bins in $\Delta E_{T,{\rm HCM}}^{\rm jet 1,2}$ are given in Table~\ref{bin_borders_tab}.
    Other details as in the
    caption to Fig. 1.
    }
  \label{fig_delet_trijet}
\end{figure}
                                %
                                %
                                %
\begin{figure}[htbp]
  
  \begin {center}
    {\fontfamily{phv}\selectfont \Huge\textbf{ZEUS}}
  \end {center}
  \begin{center}
    \begin{minipage}{\linewidth}
      \psfigadd{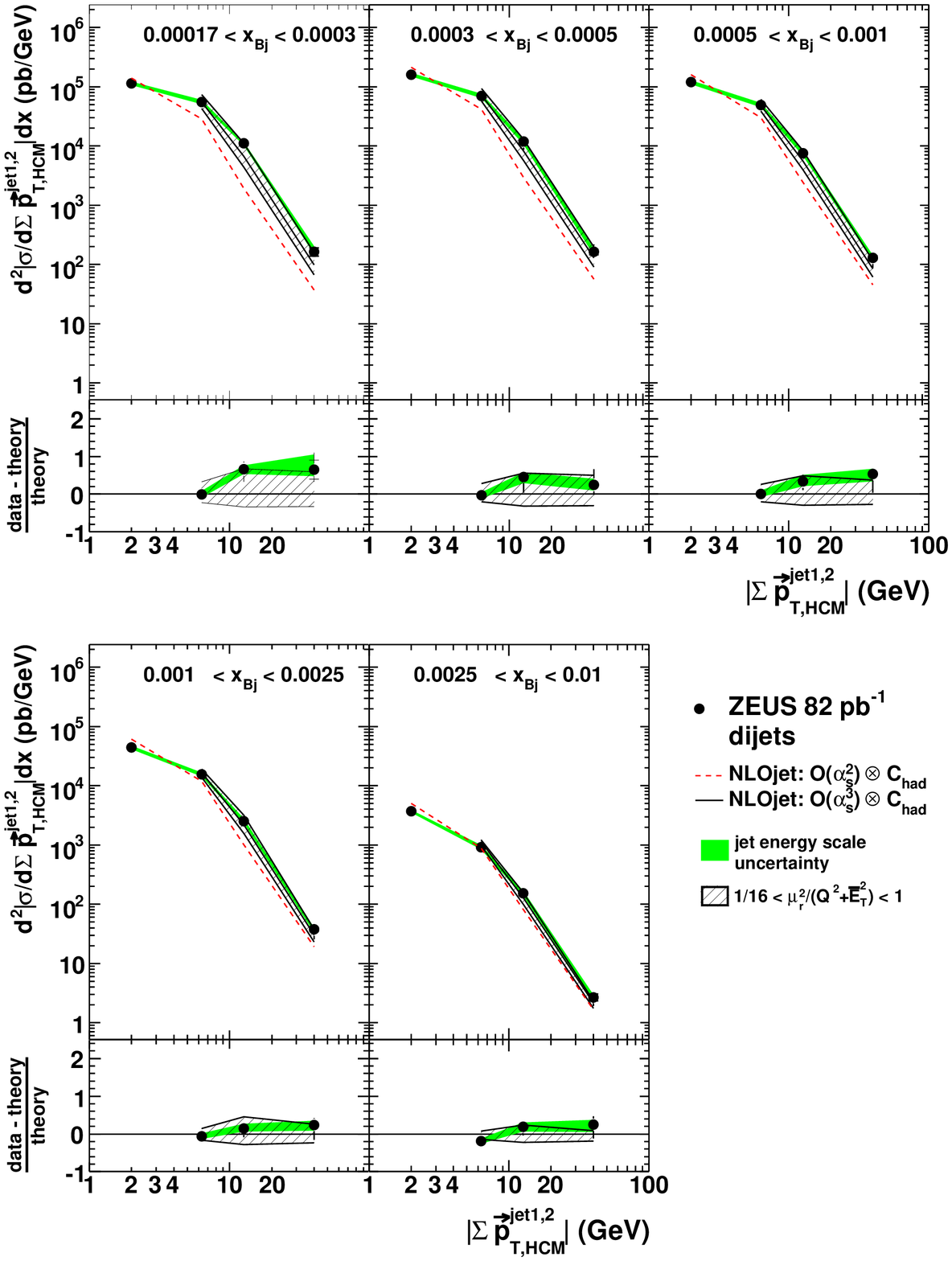}{0.9\textwidth}{1.2\textwidth}{}
    \end{minipage}
  \end{center}
  \caption{ Dijet cross sections as functions of
    $|\Sigma \vec{p}_{T,{\rm HCM}}^{~\rm jet 1,2}|$.  
    The {\sc NLOjet}
    calculations at $\mathcal{O}(\alpha_s^{2})$ $(\mathcal{O}(\alpha_s^{3}))$ are shown
    as dashed (solid) lines.
    The lower parts of the plots show the relative difference between the data and the $\mathcal{O}(\alpha_s^{3})$ predictions.  
    The boundaries for the bins in $|\Sigma \vec{p}_{T,{\rm HCM}}^{~\rm jet 1,2}|$ are given in Table~\ref{bin_borders_tab}.
    Other details as in the caption to Fig. 1.
    }
  \label{fig_sumpt_dijet}
\end{figure}
                                %
                                %
                                %
\begin{figure}[htbp]

  \begin {center}
    {\fontfamily{phv}\selectfont \Huge\textbf{ZEUS}}
  \end {center}
  \begin{center}
    \begin{minipage}{\linewidth}
      \psfigadd{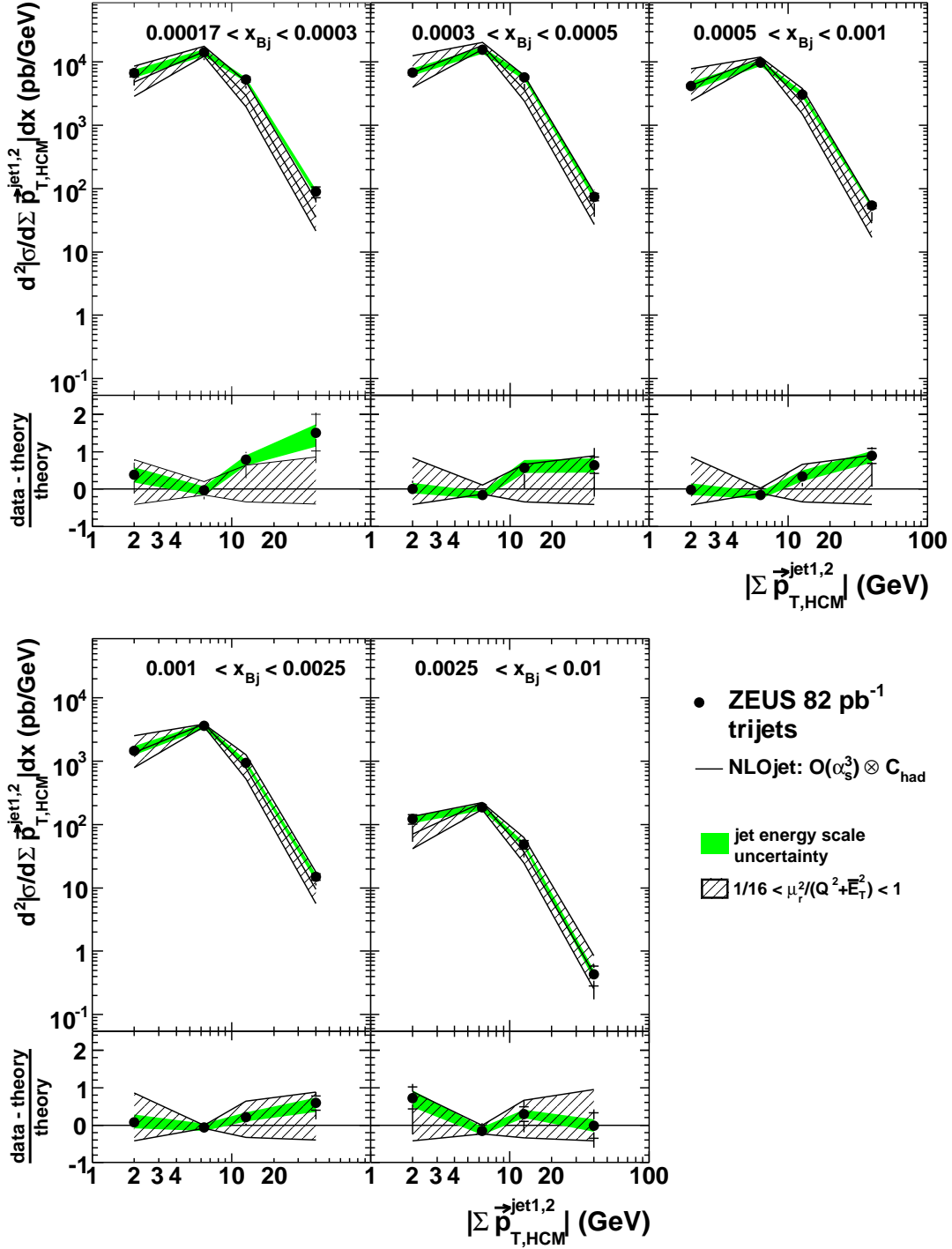}{0.9\textwidth}{1.2\textwidth}{}
    \end{minipage}
  \end{center}
  \caption{ Trijet cross sections as functions of
    $|\Sigma \vec{p}_{T,{\rm HCM}}^{~\rm jet 1,2}|$.  
    The measurements are compared to {\sc NLOjet}
    calculations at $\mathcal{O}(\alpha_s^{3})$. 
    The boundaries for the bins in $|\Sigma \vec{p}_{T,{\rm HCM}}^{~\rm jet 1,2}|$ are given in Table~\ref{bin_borders_tab}.
    Other details as in the caption to Fig. 1.
    }
  \label{fig_sumpt_trijet}
\end{figure}
                                %
                                %
                                %
\begin{figure}[htbp]
  \begin {center}
    {\fontfamily{phv}\selectfont \Huge\textbf{ZEUS}}
  \end {center}
  
  \begin{center}
    \begin{minipage}{\linewidth}
      \psfigadd{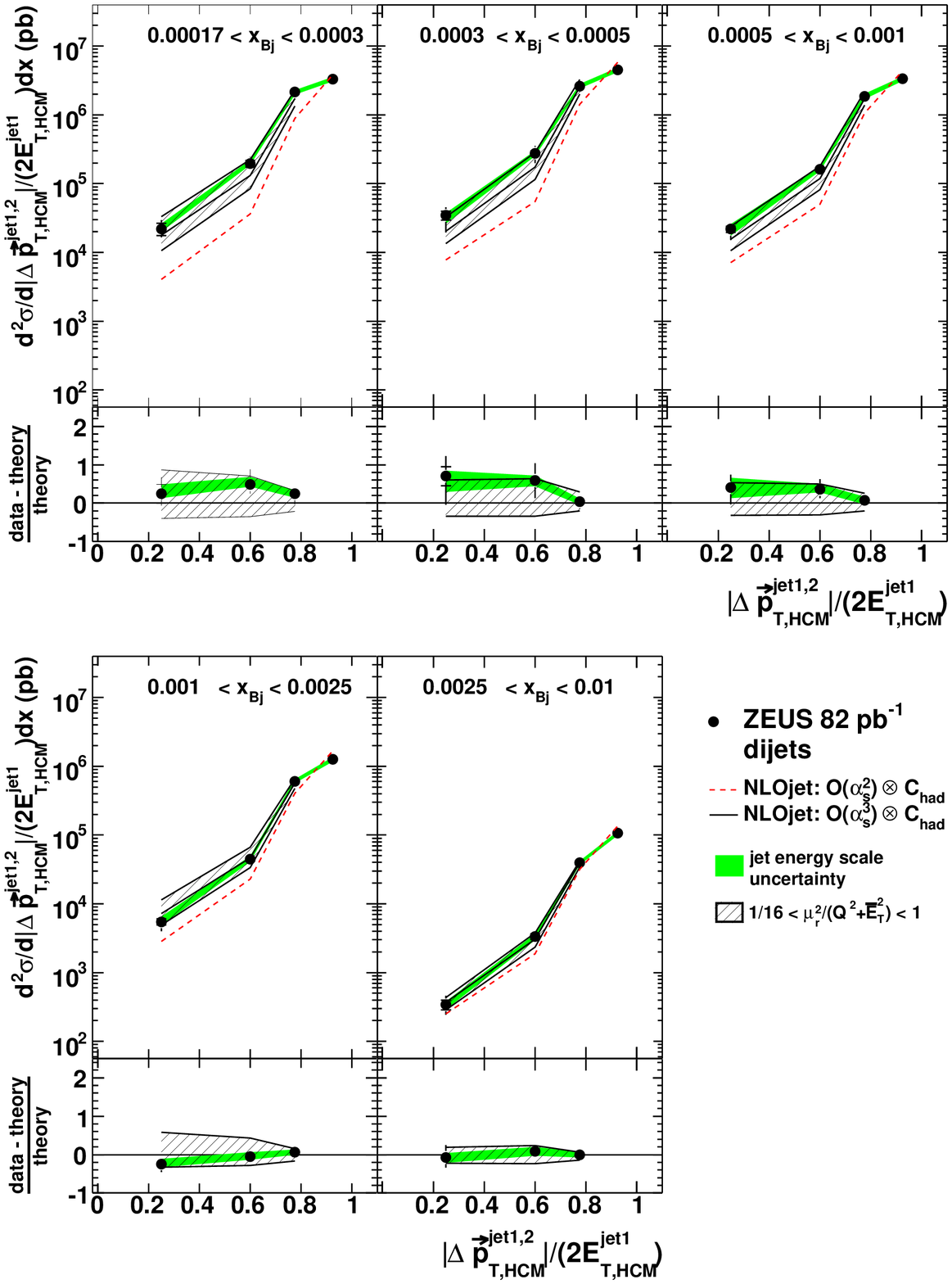}{0.9\textwidth}{1.2\textwidth}{}
    \end{minipage}
  \end{center}
  \caption{ Dijet cross sections as functions of
    $|\Delta \vec{p}_{T,{\rm HCM}}^{~\rm jet 1,2}|/(2 E_{T,{\rm HCM}}^{\rm jet 1})$.  
    The {\sc NLOjet}
    calculations at $\mathcal{O}(\alpha_s^{2})$ $(\mathcal{O}(\alpha_s^{3}))$ are shown
    as dashed (solid) lines.
    The lower parts of the plots show the relative difference between the data and the $\mathcal{O}(\alpha_s^{3})$ predictions.  
    The boundaries for the bins in $|\Delta \vec{p}_{T,{\rm HCM}}^{~\rm jet 1,2}|/(2 E_{T,{\rm HCM}}^{\rm jet 1})$ are 
    given in Table~\ref{bin_borders_tab}.
    Other details as in the caption to Fig. 1.
    }
  \label{fig_deletn_dijet}
\end{figure}
                                %
                                %
                                %
\begin{figure}[htbp]
  \begin {center}
    {\fontfamily{phv}\selectfont \Huge\textbf{ZEUS}}
  \end {center}
  
  \begin{center}
    \begin{minipage}{\linewidth}
      \psfigadd{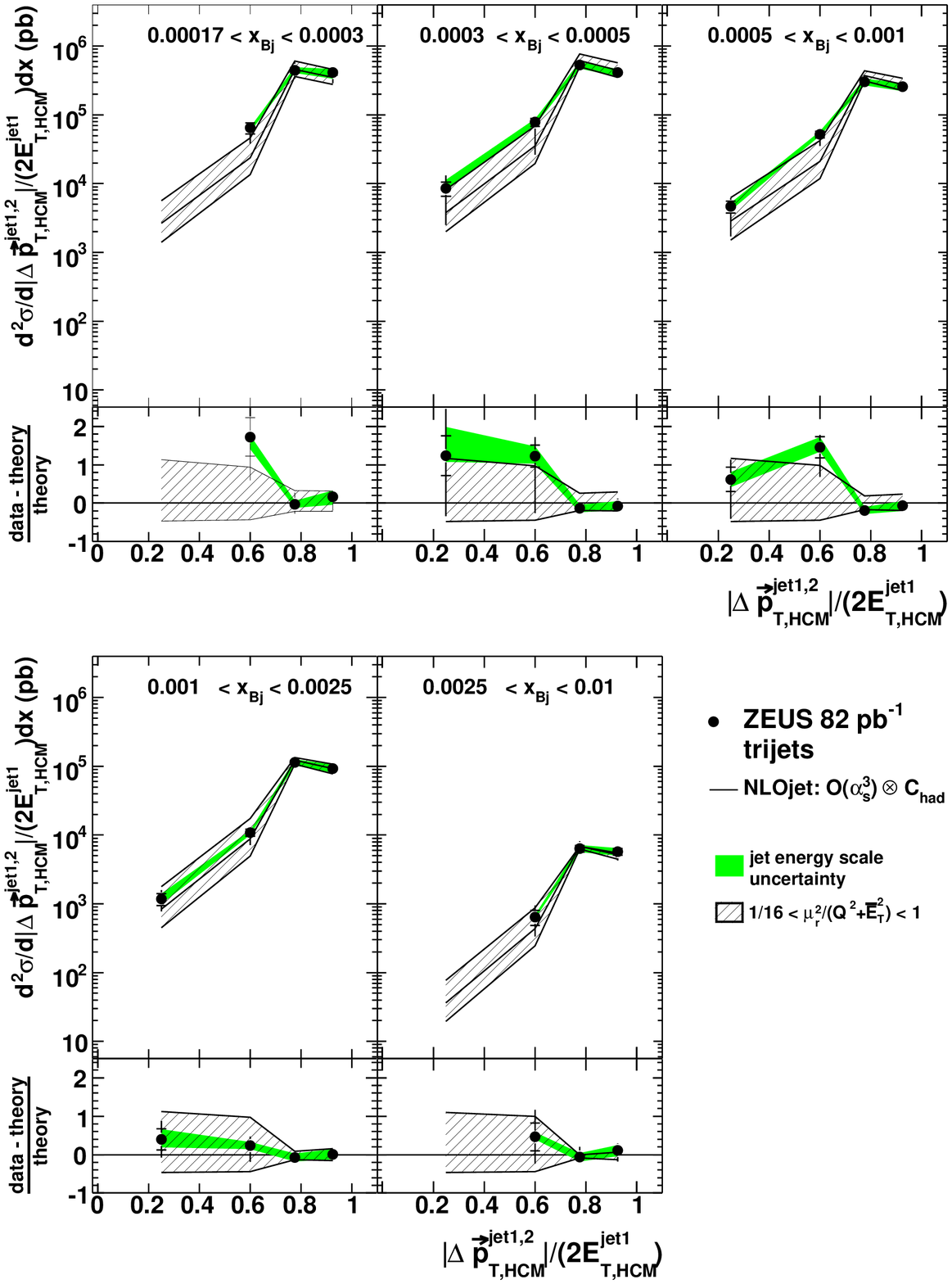}{0.9\textwidth}{1.2\textwidth}{}
    \end{minipage}
  \end{center}
  \caption{ Trijet cross sections as functions of
    $|\Delta \vec{p}_{T,{\rm HCM}}^{~\rm jet 1,2}|/(2 E_{T,{\rm HCM}}^{\rm jet 1})$.  
    The measurements are compared to {\sc NLOjet}
    calculations at $\mathcal{O}(\alpha_s^{3})$. 
    The boundaries for the bins in 
    $|\Delta \vec{p}_{T,{\rm HCM}}^{~\rm jet 1,2}|/(2 E_{T,{\rm HCM}}^{\rm jet 1})$ are given in Table~\ref{bin_borders_tab}.
    Other details as in the caption to Fig. 1.
    }
  \label{fig_deletn_trijet}
\end{figure}
                                %
                                %
\begin{figure}[htbp]

  \begin {center}
    {\fontfamily{phv}\selectfont \Huge\textbf{ZEUS}}
  \end {center}
  \begin{center}
    \begin{minipage}{\linewidth}
      \psfigadd{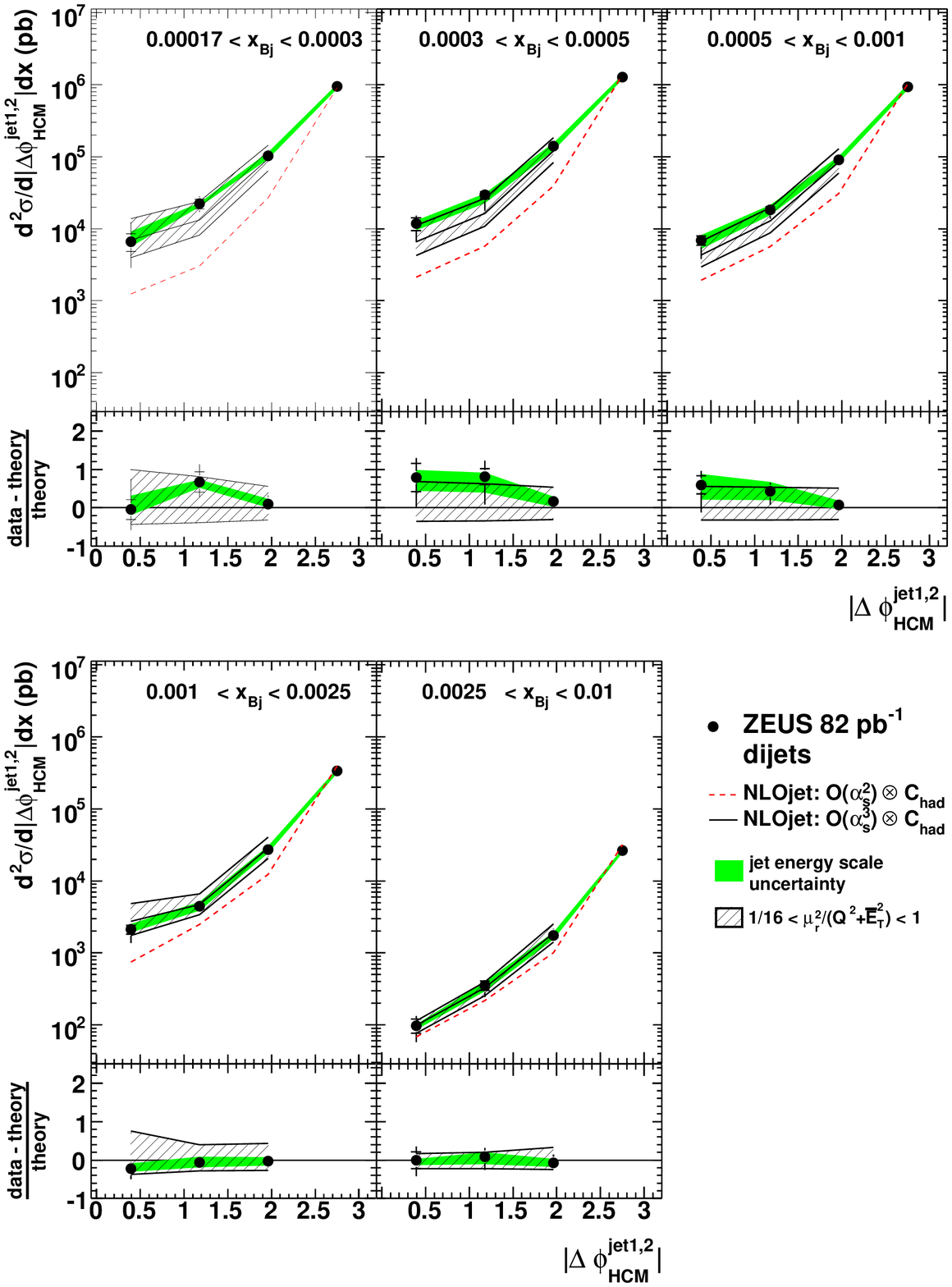}{0.9\textwidth}{1.2\textwidth}{}
    \end{minipage}
  \end{center}
  \caption{ Dijet cross sections as functions of
    $|\Delta\phi_{{\rm HCM}}^{\rm jet 1,2}|$.  
    The {\sc NLOjet}
    calculations at  $\mathcal{O}(\alpha_s^{2})$ $(\mathcal{O}(\alpha_s^{3}))$ are shown
    as dashed (solid) lines.
    The lower parts of the plots show the relative difference between the data and the $\mathcal{O}(\alpha_s^{3})$ predictions.  
    The boundaries for the bins in $|\Delta \phi_{{\rm HCM}}^{\rm jet 1,2}|$ are given in Table~\ref{bin_borders_tab}.
    Other details as in the caption to Fig. 1.
    }
  \label{fig_delphi_dijet}
\end{figure}
                                %
\begin{figure}[htbp]

  \begin {center}
    {\fontfamily{phv}\selectfont \Huge\textbf{ZEUS}}
  \end {center}
  \begin{center}
    \begin{minipage}{\linewidth}
      \psfigadd{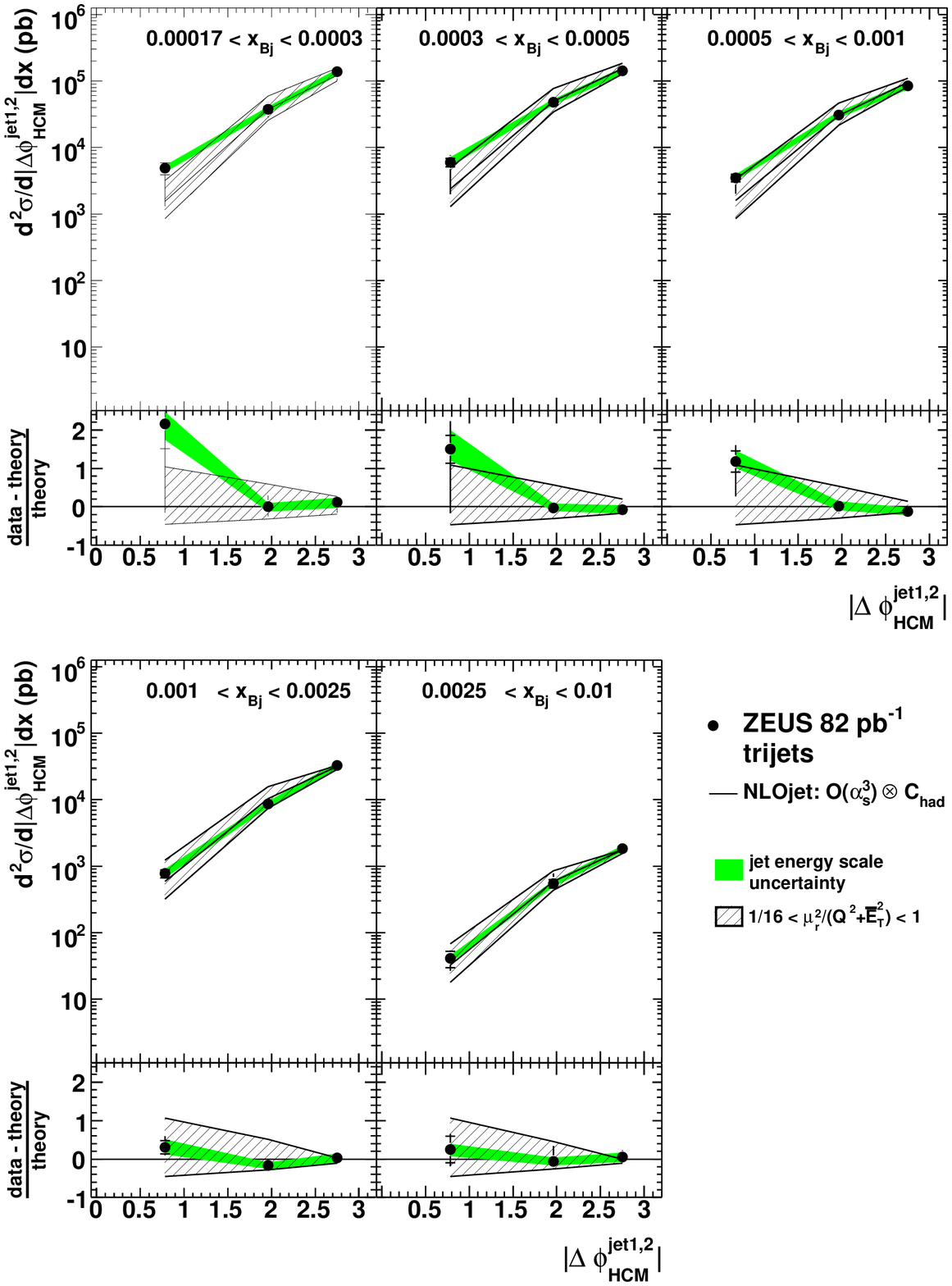}{0.9\textwidth}{1.2\textwidth}{}
    \end{minipage}
  \end{center}
  \caption{ Trijet cross sections as functions of
    $|\Delta\phi_{{\rm HCM}}^{\rm jet 1,2}|$.  
    The measurements are compared to {\sc NLOjet}
    calculations at $\mathcal{O}(\alpha_s^{3})$. 
    The boundaries for the bins in $|\Delta \phi_{{\rm HCM}}^{\rm jet 1,2}|$ are given in Table~\ref{bin_borders_tab}.
    Other details as in the caption to Fig. 1.
    }
  \label{fig_delphi_trijet}
\end{figure}
                                %
                                %
\begin{figure}[htbp]
  \begin {center}
    {\fontfamily{phv}\selectfont \Huge\textbf{ZEUS}}
  \end {center}

  \begin{center}
    \begin{minipage}{\linewidth}
      \psfigadd{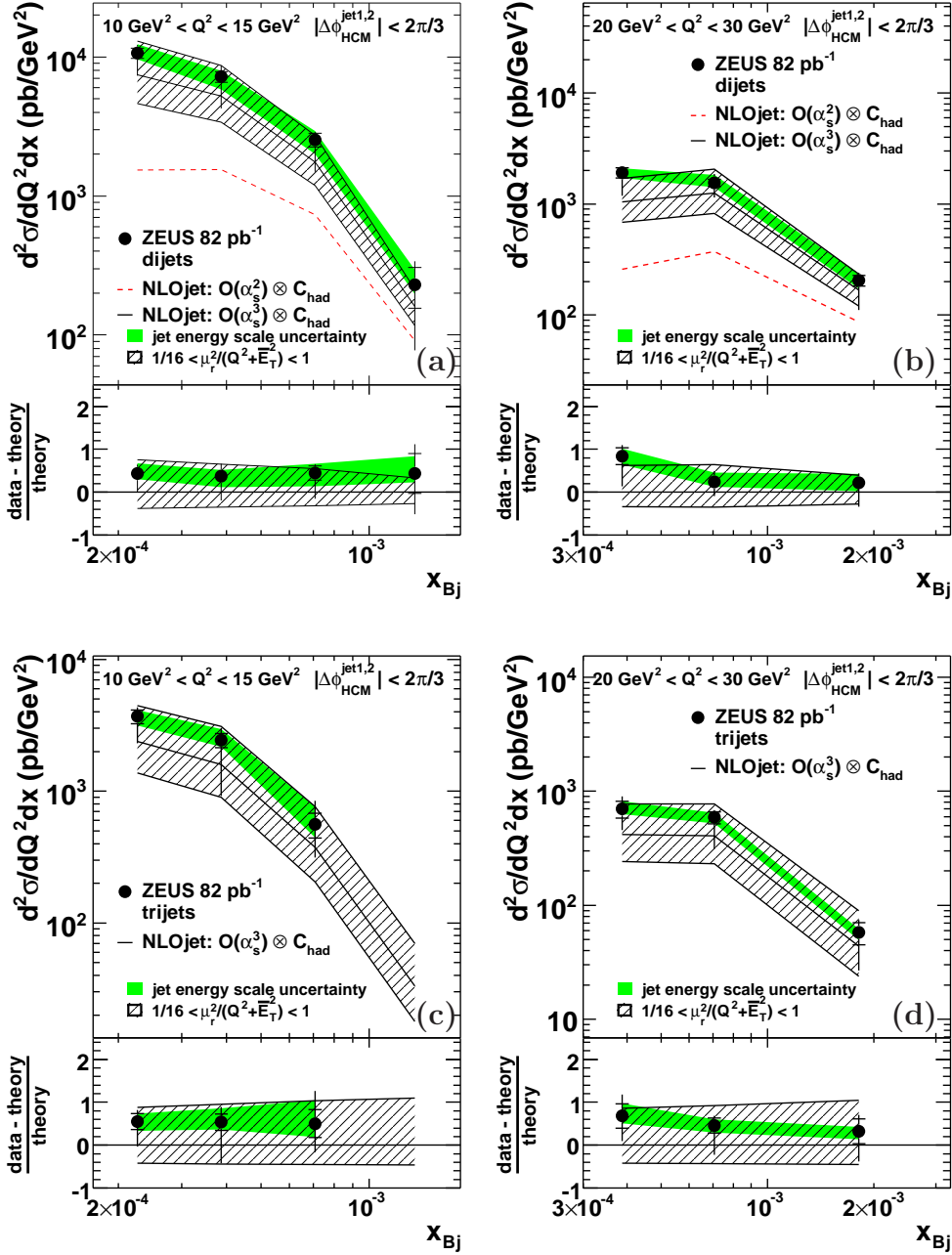}{0.8333\textwidth}{1.1111\textwidth}{%
        \Text(600,1230)[]{\bf (a)} \Text(1250,1230)[]{\bf (b)} \Text(600,340)[]{\bf (c)} \Text(1250,340)[]{\bf (d)}
        }
    \end{minipage}
  \end{center}
  \caption{ The dijet and trijet cross sections for events with
    $|\Delta\phi_{{\rm HCM}}^{\rm jet 1,2}| < 2\pi / 3$ as functions of $x_{Bj}$ in two
    different $Q^{2}$-bins.  The {\sc NLOjet}
    calculations at $\mathcal{O}(\alpha_s^{2})$ $(\mathcal{O}(\alpha_s^{3}))$ are shown
    as dashed (solid) lines. The trijet measurements are
    compared to {\sc NLOjet} calculations at $\mathcal{O}(\alpha_s^{3})$. 
    The lower parts of the plots in (a) and (b) show the relative difference between the data and the 
    $\mathcal{O}(\alpha_s^{3})$ predictions.  Other details as in the caption to Fig. 1. 
    }
  \label{fig_low_delphi}
\end{figure}

%
%
\end{document}